\documentstyle[aps,multicol,epsf]{revtex}                        
\begin{document}
\draft
\title{Quantum Phase Transitions in Superconducting Arrays \\
  with General Capacitance Matrices}

\author{Beom Jun Kim, Jeenu Kim, Sung Yong Park, and M.Y. Choi}
\address{
     Department of Physics and Center for Theoretical Physics\\
     Seoul National University\\
     Seoul 151-742, Korea}

\preprint{\today}
\maketitle
\thispagestyle{empty}

\begin{abstract}
We investigate quantum phase transitions in two-dimensional superconducting
arrays with general capacitance matrices and discrete charge states.
We use the perturbation theory together with the simulated annealing method
to obtain the zero-temperature phase diagrams, which display various 
lobe-like structures of insulating solid phases, and examine the 
possibility of supersolid phase. 
At nonzero temperatures, an effective
classical Hamiltonian is obtained through the use of the variational
method in the path-integral formalism, and the corresponding phase diagram 
is found approximately. The insulating lobes of the solid phases are shown 
to exist at sufficiently low temperatures, and 
results of Monte Carlo simulations are also presented.
\end{abstract}
\bigskip

% 74.50.+r : Proximity effects, weak links, tunneling phenomena, and Josephson
%             effects.
% 67.40.Db : Quantum statistical theory; ground state, elementary excitation.
% 05.30.Jp : Bosonic systems.

\begin{multicols}{2}

\section{Introduction}\label{sec_introduction}

Two-dimensional (2D) superconducting arrays, weakly coupled by Josephson
junctions, have attracted much attention~\cite{proceeding}. In the classical
array, where the charging energy is neglected, the relevant variable is
the phase of the superconducting order parameter at each grain, and 
the logarithmic interaction between vortices is well known to lead to
the Berezinskii-Kosterlitz-Thouless (BKT) transition~\cite{BKT}. 
Recent development of the fabrication techniques, on the other hand, allows
to make regular arrays composed of very small superconducting grains. In such
arrays the charging energy cannot be neglected any longer~\cite{geerligs}, 
and studies
considering both the self-capacitance $C_0$ and the junction capacitance $C_1$
are required. In the presence of Ohmic dissipation, the charges, which
are conjugate to the phase variables, take
continuous values and the corresponding phase diagrams have been obtained through the
use of a variational method~\cite{bjkim}. This has revealed the existence
of a low-temperature reentrant transition, which appears consistent with the 
recent quantum Monte Carlo study~\cite{rojas}. Without dissipation,
in contrast, the charges change in discrete quanta $2e$ with $-e$ being the
electron charge. In this case the phase diagrams have been investigated via the
coarse-graining approximation, which is mean-field-like in 
nature~\cite{bruder}. In general, the mean-field 
approximation does not give reliable results in two dimensions, where 
fluctuations are too strong to be neglected. It is thus desirable to study 
the 2D superconducting arrays, with general capacitance matrices
and discrete charge states, beyond the mean-field approximation.

The quantum phase transitions in the superconducting arrays have drawn much
interest also in relation to the Bose-Hubbard model, which describes strongly 
interacting bosons under the competition between the kinetic-energy and
the potential-energy effects. Here the kinetic energy and the potential
energy correspond to the Josephson energy and the charging energy in the 
superconducting array, respectively. In the presence of the on-site potential, 
the system at zero temperature exhibits a transition between
the Mott insulating phase and the superfluid phase, displaying lobe-like
structures in the resulting phase diagram~\cite{mpafisher,freericks}. 
In addition, nearest-neighbor interactions and next nearest-neighbor 
interactions produce  various phases such as the checker-board-type
solid, the striped solid, and the supersolid. In particular, the
supersolid phase, which is characterized by 
superfluidity in the presence of underlying commensurability,
has been studied extensively~\cite{meisel,otterlo}.

This paper investigates the phase transitions in 2D superconducting arrays with
emphasis on the effects of charge frustration and on the competition
between the self- and junction capacitances. 
At zero temperature, we use the perturbation expansion and the simulated 
annealing method to obtain various insulating lobes.
Here rational charge frustration introduces commensurability effects
to the array, leading to the solid phases with various charge densities. 
The detailed phase boundaries depend crucially on the ratio of the junction
to self-capacitances, $C_1/C_0$: For example, the insulating lobes become 
narrower and their central positions approach the charge frustration values, 
as $C_1/C_0$ is increased. 
Further, as the system size is increased, more and more insulating lobes
are expected to be observed in the phase diagram, suggesting the 
interesting possibility of the lobe at every rational frustration in 
the thermodynamic limit.
To check the robustness of the lobes against finite temperatures, 
we apply the Monte Carlo (MC) method to the system without the Josephson
term and find the persistence of the insulating lobes with low-order
rational values of the charge density.
We also use a variational method to obtain an effective classical
Hamiltonian describing the system at finite temperatures. 
The corresponding phase diagrams are found 
approximately in a self-consistent manner, which again  display lobes with 
the half-filled charge density. 
In the absence of charge frustration, the effective classical
Hamiltonian has only real terms, which allows us to use the MC method: We obtain
the critical strengths of the Josephson coupling and compare them with ones
obtained from the self-consistent approximation. In the self-charging limit,
the possibility of a reentrant transition is also pointed out.

This paper is organized as follows:
Section~\ref{sec_model} introduces the quantum mechanical Hamiltonian
for the superconducting arrays with general capacitance matrices and
discrete charge states. Such a superconducting array is described by quantum
mechanical conjugate variables, charges and phases, and charge
frustration is induced by applying an external voltage between the array
and the substrate. In Sec.~\ref{sec_zero}, we study the zero-temperature
phase transitions in the system via the perturbation expansion 
together with the simulated
annealing method. The phase transitions at vanishing charge density is
investigated in Sec.~\ref{subsec_0}, where the results in the nearest-neighbor 
charging limit are also compared with those of the previous works.
In Sec.~\ref{subsec_pq}, phase transitions at finite charge densities
are studied, and the nature as well as the existence of the supersolid
phase is discussed. 
Section~\ref{sec_finite} is devoted to the investigation
of the finite-temperature phase transitions by means of a variational
method and MC simulations. 
The system without the Josephson coupling is described
by only charge variables, and we present the corresponding MC results 
in Sec.~\ref{subsec_mc}, where the charge densities are 
obtained as functions of the charge frustration at various temperatures. 
The resulting step structures at zero temperature are compared with those 
obtained in Sec.~\ref{subsec_pq}. 
Whereas the phase diagrams in the presence of the Josephson coupling 
are obtained in a self-consistent manner
in Sec.~\ref{subsec_var}, the comparison with MC results reveals that the
self-consistent approximation is accurate only at sufficiently
high temperatures. Finally, a summary of the paper is given in
Sec.~\ref{sec_conclusion}.

\section{Model Hamiltonian} \label{sec_model}
We consider an $L\times L$ ($N \equiv L^2$) square array of Josephson junctions.
The extra charge $Q_i$ on the superconducting grain at site $i$ can be 
written as
\end{multicols}
\noindent\rule{0.5\textwidth}{0.1ex}\rule{0.1ex}{2ex}\hfill
\begin{eqnarray} \label{eq_Q}
Q_i & = & C_0 \Phi_i + C_1(\Phi_i - \Phi_{i+{\hat{\bf x}}})+C_1(\Phi_i - \Phi_{i-{\hat{\bf x}}})
      + C_1(\Phi_i - \Phi_{i+{\hat{\bf y}}}) + C_1(\Phi_i - \Phi_{i-{\hat{\bf y}}}) \nonumber \\
   & \equiv & \sum_j C_{ij}\Phi_j ,
\end{eqnarray}
where $C_0$ and $C_1$ are the self- and the junction capacitances, 
respectively, and
$\Phi_i$ is the electric potential of the grain $i$ with respect to
the substrate. If we apply an external voltage $\Phi_x$ between the substrate 
and the ground, Eq.~(\ref{eq_Q}) is changed to 
$Q_i = \sum_j C_{ij} \Phi_j$+$Q_x$,
with the ``gauge charge'' $Q_x \equiv C_0\Phi_x$. 
The Hamiltonian describing such a superconducting array consists
of two parts:
\begin{eqnarray} \label{eqH1}
H &=& H_0 + V \nonumber \\
  &=& \frac{1}{2} \sum_{i,j} (Q_i - Q_x)C_{ij}^{-1}(Q_j-Q_x)
     -E_J\sum_{\langle i,j\rangle}\cos(\phi_i - \phi_j ),
\end{eqnarray}
\hfill\raisebox{-1.9ex}{\rule{0.1ex}{2ex}}\rule{0.5\textwidth}{0.1ex}
\begin{multicols}{2}
\noindent
where $\phi_i$ represents the phase of the superconducting order
parameter at site $i$. Note that the charge $Q_i$ and the phase
$\phi_i$ are quantum mechanical conjugate variables satisfying
the commutation relation $[\phi_i, Q_i] = 2ei$ ($e>0$). 
In the charging energy term $H_0$, the charges at sites $i$ and $j$ 
interact via the inverse of the capacitance matrix $C_{ij}^{-1}$
whereas the summation in the Josephson energy term $V$ is to be done
over all the nearest neighboring pairs. 
When the lattice constant of the array approaches zero,
we can write Eq.~(\ref{eq_Q}) in the continuum form $Q({\bf r}) = C_0\Phi({\bf r})
-C_1\nabla^2 \Phi({\bf r})$ and obtain $\Phi({\bf r})$ due to a single 
positive charge
at the origin~\cite{fazio}: $\Phi({\bf r}) \propto K_0(r/\Lambda)$ with the
modified Bessel function $K_0$, where the screening length $\Lambda$ ($\equiv 
\sqrt{C_1/C_0}$) measures the interaction range between charges. 
The above Hamiltonian
in Eq.~(\ref{eqH1}) is closely related to the Bose-Hubbard Hamiltonian and 
to the spin-1/2 $XXZ$ Hamiltonian~\cite{otterlo}.
The charging energy term in Eq.~(\ref{eqH1}) drives the system to
arrange its discrete charges such that the average charge per site is as
close to $Q_x$ as possible, and the Josephson energy 
corresponds to the hopping energy in the Bose-Hubbard model~\cite{mcha}.
Consequently, if the charging energy is sufficiently larger than the
Josephson energy, hopping is suppressed and all the charges are
arranged to form a periodic lattice, leading to the insulating phase 
at zero temperature. In particular, when 
$E_J = 0$, the energy eigenstate can be chosen also as a 
simultaneous eigenstate of the charge operator since $[Q_i,H]=0$.
The uncertainty relation between the charge and the phase then
asserts that phase coherence and consequently, superconductivity
cannot be attained in this limit. 
In the opposite limit of the strong Josephson coupling, all the charges are
in the extended state owing to  the dominant hopping term, and 
the superconducting phase is favored. 
For convenience, we write $q_i \equiv Q_i/2e$, $q_x \equiv Q_x/2e$,
$E_0 \equiv e^2/2C_0$, and ${\widetilde C}_{ij} \equiv
C_{ij}/C_0$, and obtain Eq.~(\ref{eqH1}) in the form
\end{multicols}
\noindent\rule{0.5\textwidth}{0.1ex}\rule{0.1ex}{2ex}\hfill
\begin{equation} \label{eqH2}
  H  = H_0 + V = 4E_0\sum_{i,j} (q_i - q_x){\widetilde C}_{ij}^{-1}(q_j-q_x)
      -E_J\sum_{\langle i,j \rangle}\cos (\phi_i - \phi_j) , 
\end{equation}
\hfill\raisebox{-1.9ex}{\rule{0.1ex}{2ex}}\rule{0.5\textwidth}{0.1ex}
\begin{multicols}{2}
\noindent
where the inverse of the dimensionless capacitance matrix is given by
\begin{equation}
{\widetilde C}_{ij}^{-1} = \frac{1}{N}\sum_{\bf k}
    \frac{e^{i{\bf k}\cdot ({\bf x}_i - {\bf x}_j)}}
         {1 + 4C_1/C_0 - 2(C_1/C_0)(\cos k_x + \cos k_y)} .
\end{equation}

\section{Zero-temperature phase diagrams} \label{sec_zero}
 
\begin{figure}
\centerline{\epsfxsize=8.0cm \epsfbox{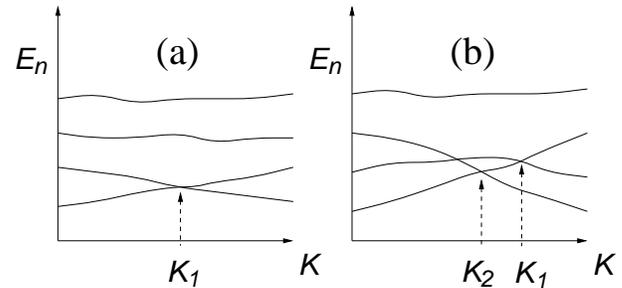}}
\caption{Schematic diagram of the energy levels $E_n$ versus the control parameter
$K$.
In (a) the crossing point $K_1$ of the ground state and the first excited state 
corresponds to the transition point. In (b), on the other hand, the 
true transition point is given by $K_2$, at which the ground state meets
the second excited state. }
\label{fig_sch}
\end{figure}

In general the state which minimizes the energy determines the zero-temperature 
phase of the system. Suppose that we have obtained the
energy levels of the system which correspond to the ground state,
the first excited state, etc. If we introduce a small change
in the control parameters, the energy levels shift by a small amount,
and it is possible that the ground state and the first excited state
cross each other as we pass through a set of values of the control parameters;
this ``level crossing'' leads to a zero-temperature phase transition 
in the parameter space~\cite{book}.  
It should be noted here that consideration of
only the two low-lying states, the ground state and the first excited
one, may not be sufficient. Figure~\ref{fig_sch}
shows schematic dependence of the energy levels $E_n$ on
the control parameter $K$ in two possible cases. In Fig.~\ref{fig_sch}~(a),
the ground state and the first excited state cross at the critical 
value $K_1$ of the control parameter, which corresponds to the true
transition point. In Fig.~\ref{fig_sch}~(b), on the other hand, the true
transition point is given by $K_2$, where the ground state and the
second excited one cross each other.  In this case consideration of
only the two low-lying states would lead to $K_1$, which is not the true
transition point. To find the true transition point, we thus need to consider
all the energy levels, which is not possible in practice.
Despite this, considering the level crossing of the ground state and the
first excited state is very useful and convenient to obtain the upper limit
of the true transition point.

In terms of charges the system can be either in the incommensurate
conducting phase or in the commensurate insulating phase; in the latter 
charges are expected to form a superlattice which covers periodically the
whole system.
If the commensurate state has energy lower/higher than that of the 
incommensurate state for given values of parameters, the system should be
in the insulating/conducting phase.
(Here $E_J/E_0$, $q_x$, and $C_1/C_0$ constitute the control parameters.)
Accordingly, for given total number of charges, 
it is needed to find the charge configurations of the commensurate state
and of the low-lying incommensurate states.
The symmetry of the Hamiltonian under the transformations
\begin{equation}
\left\{
\begin{array}{lll}
q_x & \rightarrow & q_x +1 , \nonumber  \\
q_x & \rightarrow & -q_x  , 
\end{array}
\right.
\end{equation}
allows us to consider $q_x$ only in the range $0 \le q_x < 1/2$. 
In this restricted range $0 \le q_x<1/2$, the commensurate states 
of the system may be classified in terms of the average number
of charges, $\langle q_i \rangle $ $\equiv \sum_i q_i/N$.
For $\langle q_i \rangle  = p/q$ with relatively prime integers
$p$ and $q$, the charges are expected to form a $q\times q$ 
superlattice, on the unit cell of which $pq$ charges, each with the
unit charge, are located.

\subsection{Phase transitions for $\langle q_i \rangle = 0$} \label{subsec_0}

We first investigate the phase transitions in the system with  
$\langle q_i \rangle = 0$. The energy of the 
$\langle q_i \rangle = 0$ commensurate state is computed via
the second-order perturbation expansion with the Josephson term as
a perturbation in Eq.~(\ref{eqH2}), and is written as 
\begin{equation}
E_{(c)}(\langle q_i \rangle = 0) = E_{(c)}^{(0)}(0) + 
      E_{(c)}^{(1)}(0) + E_{(c)}^{(2)}(0) + O(E_J^3).
\end{equation}
Without the perturbation, the Hamiltonian of the system contains only
charge operators and yields the eigenfunction
\end{multicols}
\noindent\rule{0.5\textwidth}{0.1ex}\rule{0.1ex}{2ex}\hfill
\begin{equation}
\langle \mbox{\boldmath{$\phi$}} | {\bf q} \rangle \equiv 
\langle \phi_1, \phi_2, \phi_3, \cdots , \phi_N | 
       q_1, q_2, q_3, \cdots , q_N \rangle = 
    \left( \frac{1}{\sqrt{2\pi}} \right)^N
    \exp\left(i\sum_iq_i \phi_i\right),
\end{equation}
%\hfill\raisebox{-1.9ex}{\rule{0.1ex}{2ex}}\rule{0.5\textwidth}{0.1ex}
\begin{multicols}{2}[]
\noindent
which is simply the  plane-wave state of
the usual free particle system, with the interpretation of $\phi_i$ and
$q_i$ as the position and the momentum, respectively.
For the commensurate state with $\langle q_i  \rangle
= 0$, all the charge eigenvalues $q_i$'s are zero and 
it is easy to compute
the zeroth-order contribution $E_{(c)}^{(0)}(0)$  
and the first-order contribution $E_{(c)}^{(1)}(0)$: 
$E_{(c)}^{(0)}(0)$  = $4E_0 q_x^2$$\sum_{i,j}{\widetilde C}_{ij}^{-1}$ 
              = $4E_0 q_x^2N $ and 
$E_{(c)}^{(1)}(0)$ = $\langle {\bf q} = 0| V | 
             {\bf q} = 0 \rangle$ 
                 = $(1/2\pi)^N\int 
             \left(\prod_i d\phi_i\right)$
                     $\sum_{\langle i,j\rangle}\cos(\phi_i - \phi_j) =  0.$
The second-order contribution is given by 
\begin{equation} \label{eq_ec2}
E_c^{(2)}(0) = \sum_{ {\bf q}' \ne 0 } \frac{ |\langle {\bf q}' |V|
     {\bf q}=0 \rangle |^2 }{ E^{(0)}( {\bf q} = 0) - E^{(0)}({\bf q}')} ,
\end{equation}
where $E^{(0)}( {\bf q} )$ is the zeroth-order energy for the charge
configuration ${\bf q}$, and
\end{multicols}
\noindent\rule{0.5\textwidth}{0.1ex}\rule{0.1ex}{2ex}\hfill
\begin{eqnarray} \label{eq_v}
\langle {\bf q}' |V| {\bf q}=0 \rangle &=& 
   \langle {\bf q}' | -E_J\sum_{\langle i,j\rangle}\cos(\phi_i - \phi_j) |
    {\bf q}=0 \rangle \nonumber \\
&=& - \frac{E_J}{ 2 (2\pi)^N } \sum_{\langle i,j\rangle}\int\prod_k d\phi_k
    \left[ e^{ i(\phi_i - \phi_j - \sum_k q'_k\phi_k) } +
    e^{ i(\phi_j - \phi_i - \sum_k q'_k\phi_k) } \right]
\nonumber \\
&=& -\frac{E_J}{2}\sum_{\langle i,j\rangle}\left[\prod_{k\ne i,j} 
 \delta(q'_k,0)\right] \left[\delta(q'_i,1) \delta(q'_j,-1) +
    \delta(q'_j,1) \delta(q'_i,-1) \right] .
\end{eqnarray}
\hfill\raisebox{-1.9ex}{\rule{0.1ex}{2ex}}\rule{0.5\textwidth}{0.1ex}
\begin{multicols}{2}[]
\noindent
Inserting Eq.~(\ref{eq_v}) into Eq.~(\ref{eq_ec2}), we obtain the 
energy of the $\langle q_i \rangle = 0$
commensurate state to the 2nd order in $E_J$: 
\begin{equation} \label{eq_Ec0}
E_{(c)}(0) \approx 4E_0Nq_x^2 - \frac{E_J^2 N}{8E_0 \left(
 {\widetilde C}_{00}^{-1} - {\widetilde C}_{10}^{-1} \right) }.
\end{equation}

We next consider the incommensurate state with $\langle q_i \rangle =0$.
Among the various excitations present in the system with the constraint
$\langle q_i \rangle = 0$, the lowest excitation is expected to be
point-like and we consider two types of excitations: One is the creation of 
a single charge (SC) at one site~\cite{freericks}, and the other is the
creation of a charge dipole (CD) at one bond [$q_i=1$, $q_j=-1$ with $(i,j)$ being the
nearest-neighboring pair]. Both excitations do not change
the value of $\langle q_i \rangle$ in the thermodynamic limit
($N\rightarrow \infty$). In the self-charging limit ($C_1 = 0$),
the interaction matrix $C_{ij}^{-1}$ takes
a diagonal form and the first excited state should be of the SC type,
since the CD type excitation requires the creation energy of two charges.
On the other hand, in the nearest-neighbor (NN) charging limit ($C_0 = 0$),
only the excitations which does not change the total
number of charges are allowed and the first excited state is expected to be
of the CD type.
Accordingly, as the value of $C_1/C_0$ is increased, the first excited state
should change from the SC type to the CD type at a critical
value of $C_1/C_0$.  
In particular, the system in the NN charging limit displays the
charge-anticharge unbinding transition~\cite{fazio,zant,kanda}, the BKT
nature of which is manifested by the square-root cusp in the
resistance~\cite{kanda}. This charge BKT transition is suppressed by the
presence of the self-capacitance, making 
the screening length finite, and it seems likely  
that the change of the first excited state according to the value of 
$C_1/C_0$ in turn leads to the change in the nature of the transition.

\begin{figure}
\centerline{\epsfxsize=10.0cm \epsfbox{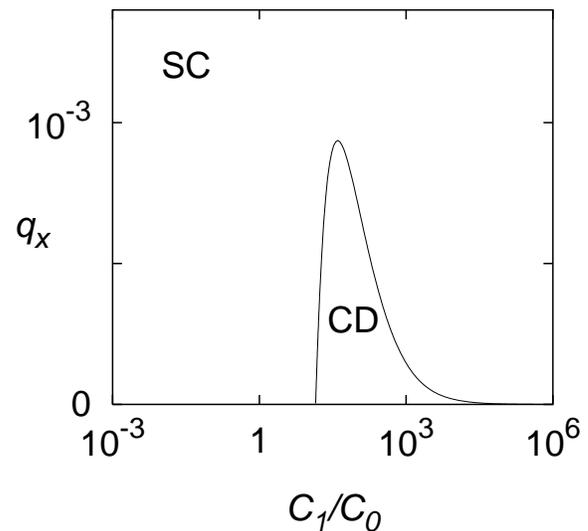}}
\vskip 0.5cm
\caption{Boundary for the first excited state when $\langle q_i \rangle =0$.
In the region enclosed by  the curve the creation of a charge dipole (CD) at 
one bond gives the lowest excitation, and the creation of a single
charge (SC) at one site is the lowest excitation in the region outside 
the curve. For $C_1/C_0 < 14.116$ the lowest excitation is SC irrespective
of $q_x$, and in the nearest-neighbor charging limit ($C_0=0$) the lowest
excitation is CD for $q_x=0$ and SC for $q_x \ne 0$, respectively.} 
\label{fig_1st}
\end{figure}

Between the above two types of excitations, the lowest
one is determined from the comparison of the zeroth-order energies of 
both types:
If $\Delta E$ given by
\end{multicols}
\noindent\rule{0.5\textwidth}{0.1ex}\rule{0.1ex}{2ex}\hfill
\begin{eqnarray}
\Delta E &\equiv& E^{(0)}( \mbox{SC} ) - 
   E^{(0)}( \mbox{CD}) \nonumber \\
&=& 4E_0\left( {\widetilde C}_{00}^{-1} - 2q_x + Nq_x^2 \right) -
    4E_0\left( {\widetilde C}_{00}^{-1} - {\widetilde C}_{01}^{-1}  
               -  {\widetilde C}_{10}^{-1} +  {\widetilde C}_{11}^{-1} 
              + Nq_x^2 \right) \nonumber \\
&=& 4E_0\left( 2{\widetilde C}_{10}^{-1} -  {\widetilde C}_{00}^{-1} 
          -2q_x \right)
\end{eqnarray}
is negative/positive, the creation of a single charge/charge dipole 
is the lowest excitation.
Figure~\ref{fig_1st} shows the boundary for $\Delta E = 0$ in the
limit of $N \rightarrow \infty$.
It is found that for $C_1/C_0 < 14.116$ the lowest excitation is of the
SC type regardless of the value of $q_x$, and that  the region 
of the CD type excitation (the region below the curve in Fig.~\ref{fig_1st}) 
is quite small, allowed only for   
$q_x < 9.4\times 10^{-4}$. Therefore we below focus on the SC type excitation
for $C_0 \ne 0$ and $q_x > 0$,  and then consider the CD type excitation
only in the NN charging limit ($C_0 =0$) with $q_x = 0$. 

\begin{figure}
\centerline{\epsfxsize=8.0cm \epsfbox{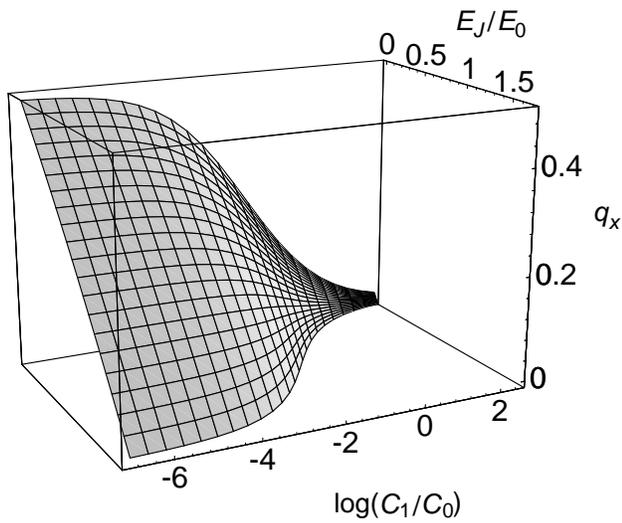}}
\vskip 0.5cm
\caption{Zero-temperature phase diagram for 
$\langle q_i \rangle = 0$.
The system is insulating in the region below the surface, which 
shrinks monotonically as $C_1/C_0$ is increased. } 
\label{q0_bound}
\end{figure}

We  write the energy of the first excited state of the SC type as 
\begin{equation}
E_{(ic)}(\langle q_i \rangle = 0 )  = 
E_{(ic)}^{(0)}(0) + E_{(ic)}^{(1)}(0) + E_{(ic)}^{(2)}(0) + O(E_J^3),
\end{equation}
and the eigenstate as 
\begin{equation}\label{eqql}
|{\bf q}^{(l)}\rangle = |q_1^{(l)},q_2^{(l)},\cdots ,q_N^{(l)}\rangle 
\mbox{\ \ \ with \ \ \ } 
\left\{
\begin{array}{ll}
q_i^{(l)} = 1 & \mbox {\ \ \ \ \ for $i=l$,}  \\
q_i^{(l)} = 0 & \mbox {\ \ \ \ \ otherwise,} 
\end{array}
\right.
\end{equation}
which is expected to be a superconducting state since the extra charge at 
site $l$ can hop to any site without energy cost.
Inserting Eq.~(\ref{eqql}) to $H_0$ in Eq.~(\ref{eqH2}), 
we obtain the zeroth-order contribution:
$E_{(ic)}^{(0)}(0)$ = $4E_0( {\widetilde C}_{00}^{-1} - 2q_x + Nq_x^2)$,
where we have used the translational symmetry ${\widetilde C}_{ll}^{-1} =
{\widetilde C}_{00}^{-1}$ and $\sum_j {\widetilde C}_{ij}^{-1} = 1$. 
Since the above excited state in Eq.~(\ref{eqql}) has an $N$-fold 
degeneracy ($l$ can take values $1,2, \cdots , N$), 
we use the degenerate perturbation theory to calculate
the first-order contribution $E_{(ic)}^{(1)}(0)$. 
We thus need to diagonalize the matrix $V$ whose components are
given by
\begin{eqnarray}
V_{lm} &\equiv& \langle {\bf q}^{(l)} |V|{\bf q}^{(m)}\rangle  \nonumber \\
       &=& \left\{ \begin{array}{cl} 
                  -E_J/2  &  \ \ \ \mbox{for $(l,m)$ nearest neighboring
                             pair, } \\ 
                   0        &  \ \ \ \mbox{otherwise.}
                 \end{array} \right.
\end{eqnarray}
For that purpose, it is convenient to use the
Fourier transform:
\begin{equation} 
V_{lm} = \frac{1}{N}\sum_{\bf k} 
              e^{i{\bf k}\cdot ({\bf x}_l - {\bf x}_m)}
             {\widetilde V}_{\bf k} ,  
\end{equation}
which gives straightforwardly the eigenvalues 
${\widetilde V}_{\bf k} = -E_J (\cos k_x + \cos k_y)$.
The first-order contribution is then given by the lowest eigenvalue:
$E_{(ic)}^{(1)}(0)= {\widetilde V}_{{\bf k} = 0} = -2E_J$. For this
lowest-lying excited state with ${\bf k} = 0$, it is also straightforward
to obtain the second-order
contribution, which leads to the energy of the incommensurate state
\begin{eqnarray} \label{eqEic0}
E_{(ic)}(0) &\approx& 4E_0(Nq_x^2 - 2q_x + {\widetilde C}_{00}^{-1}) - 2E_J
  +\frac{E_J^2}{32E_0}{\sum_{(i,j)}}^\prime \frac{1}{ {\widetilde C}_{j0}^{-1}
  -{\widetilde C}_{i0}^{-1} -{\widetilde C}_{00}^{-1} +
  {\widetilde C}_{10}^{-1}} \nonumber \\
 & & + \frac{E_J^2}{8E_0}\left( \frac{2}{ {\widetilde C}_{10}^{-1} -  
      {\widetilde C}_{\hat{\bf x}+\hat{\bf y},0}^{-1} - {\widetilde C}_{00}^{-1} 
     +{\widetilde C}_{10}^{-1}} 
     + \frac{1}{ {\widetilde C}_{10}^{-1} -  
      {\widetilde C}_{2\hat{\bf x},0}^{-1} - {\widetilde C}_{00}^{-1} 
     +{\widetilde C}_{10}^{-1}}\right).
\end{eqnarray}
Here $\sum_{(i,j)}^\prime$ denotes the summation over $i$ and its four
nearest-neighbors $j$ with the restriction $j \neq 0$.
The zero-temperature transition occurs when the energy of the commensurate
state given by Eq.~(\ref{eq_Ec0}) and that of the incommensurate state given
by Eq.~(\ref{eqEic0}) becomes equal. 
This reveals the zero-temperature phase transition in the system with
$\langle q_i \rangle = 0$, as $q_x$ is varied.
Namely, the system becomes superconducting as $q_x$ is increased beyond
the critical value $q_x^c$, which is given by
\begin{eqnarray}
q_x^c &\equiv& \frac{1}{2}\left( {\widetilde C}_{00}^{-1} - \frac{E_J}{2E_0} 
 \right) + \frac{1}{256}\left( \frac{E_J}{E_0} \right)^2 \left( \frac{4N}
               {{\widetilde C}_{00}^{-1} -{\widetilde C}_{10}^{-1}} 
   +\sum_{(i,j), j\ne 0 } 
       \frac{1}{ {\widetilde C}_{j0}^{-1}
      -{\widetilde C}_{i0}^{-1} -{\widetilde C}_{00}^{-1} +
       {\widetilde C}_{10}^{-1}} \right) \nonumber \\
 & & + \frac{1}{64}\left(\frac{E_J}{E_0}\right)^2 \left( \frac{2}{ {\widetilde C}_{10}^{-1} -  
      {\widetilde C}_{\hat{\bf x}+\hat{\bf y},0}^{-1} - {\widetilde C}_{00}^{-1} 
     +{\widetilde C}_{10}^{-1}} 
     + \frac{1}{ {\widetilde C}_{10}^{-1} -  
      {\widetilde C}_{2\hat{\bf x},0}^{-1} - {\widetilde C}_{00}^{-1} 
     +{\widetilde C}_{10}^{-1}}\right).
\end{eqnarray}
The resulting phase boundary between the insulating phase and the
superconducting phase is shown in Fig.~\ref{q0_bound}, obtained
for a $24\times 24$ array.
The junction capacitance here tends to enhance superconductivity, which is
in accord with the result in Ref.~\onlinecite{bjkim}.
It is also obvious that as
the system approaches the NN charging limit, the 
insulating region shrinks to zero; {\it this apparently cures the
well-known artifact of the coarse-graining approximation},
where the system remains in the insulating phase for arbitrarily large 
values of $E_J$ in the NN charging limit~\cite{bruder}.

\begin{figure}
\centerline{\epsfxsize=10.0cm \epsfbox{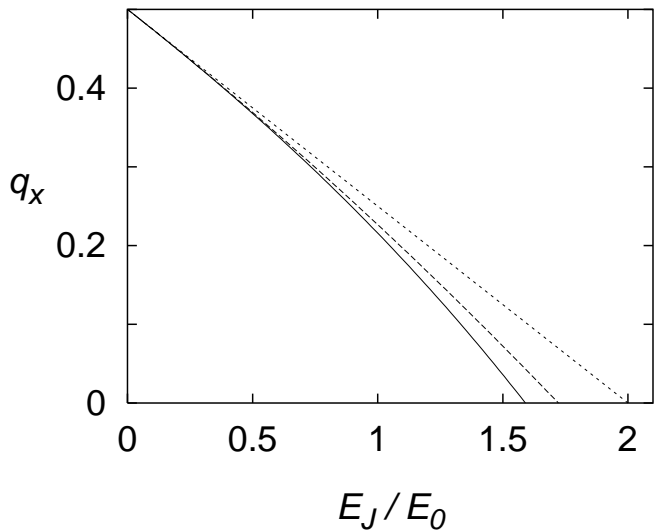}}
\vskip 0.5cm
\caption{Zero-temperature phase boundaries for $\langle q_i \rangle = 0$
in the self-charging limit,  computed via the perturbation 
expansion up to $O({E_J}^n)$. The dotted line, the dashed line, and the solid
line correspond to the first-order ($n=1$), the second-order ($n=2$,
and the third-order ($n=3$) calculations, respectively. }
\label{3rd_bound}
\end{figure}

To check the validity of the second-order perturbation expansion, we have
computed the third-order contribution in the self-charging limit,
and obtain the critical value $q_x^c$:
\begin{equation} \label{eq_3rd}
q_x^c(C_1=0) = \frac{1}{2} - \frac{1}{4}\left(\frac{E_J}{E_0}\right)
               -\frac{3}{128}\left(\frac{E_J}{E_0}\right)^2
               -\frac{11}{1024}\left(\frac{E_J}{E_0}\right)^3.
\end{equation}
\begin{multicols}{2}[]
\noindent
The three curves in Fig.~\ref{3rd_bound} represent
the phase boundaries obtained from
Eq.~(\ref{eq_3rd}) up to the first-, the second-, and the third-order 
contributions, respectively. It is shown that the third-order contribution
does not change significantly the boundary obtained from the second-order calculation,
inducing 8.3\% of the difference in the 
critical value of $E_J/E_0$ at $q_x = 0$. 
                
We next consider the NN charging limit in the absence of charge
frustration ($q_x = 0$). In this case, Eq.~(\ref{eq_Ec0}) gives
the energy of the commensurate state in the form 
\begin{equation}
E_{(c)} \approx -\frac{E_J^2 N}{8E_1\left( \bar{C}_{00}^{-1}
  -{\bar C}_{10}^{-1}\right)} ,
\end{equation}
where $E_1 \equiv e^2/2C_1$ and 
\begin{equation} \label{eq_Cbar}
{\bar C}_{ij}^{-1} \equiv \frac{1}{N}\sum_{\bf k}
    \frac{e^{i{\bf k}\cdot ({\bf x}_i - {\bf x}_j)}}
         {4 - 2(\cos k_x + \cos k_y)} .
\end{equation}
In the thermodynamic limit, we replace the discrete sum
by the integral and obtain 
\end{multicols}
\noindent\rule{0.5\textwidth}{0.1ex}\rule{0.1ex}{2ex}\hfill
\begin{eqnarray}
{\bar C}_{00}^{-1} - {\bar C}_{10}^{-1} &=&
  \frac{1}{ (2\pi)^2} \int_{-\pi}^\pi dk_x \int_{-\pi}^\pi dk_y
   \frac{ 1 - \cos k_x }{ 4 - 2( \cos k_x + \cos k_y) } \nonumber \\
&=& \frac{1}{2\pi}\int_0^\pi dk_x \sqrt{\frac{1-\cos k_x}{3-\cos k_x}}
  = \frac{1}{4} ,
\end{eqnarray}
\hfill\raisebox{-1.9ex}{\rule{0.1ex}{2ex}}\rule{0.5\textwidth}{0.1ex}
\begin{multicols}{2}[]
\noindent
leading to $E_{(c)} \approx -E_J^2N/2E_1$.
The first-excited incommensurate state in the NN charging limit is of the
CD type (see Fig.~\ref{fig_1st}) with the zeroth-order energy
$E_{(ic)}^{(0)}$=$8E_1( {\bar C}_{00}^{-1}$-${\bar C}_{10}^{-1})$
=$2E_1$ and the null first-order energy $E_{(ic)}^{(1)} = 0$.
It is complicated but straightforward to calculate 
$E_{(ic)}^{(2)}$ from the second-order degenerate perturbation theory, which
leads to a $4N \times 4N$ matrix. The Fourier transform then yields 
$N$ $4\times 4$ Hermitian matrices, each of which can be diagonalized.
The lowest eigenvalue then gives
the energy of the CD type incommensurate state, up to the second order:
\begin{equation}
E_{(ic)} \approx 2E_1 - \frac{E_J^2}{8E_1}\left( 4N + 62.000 \right) ,
\end{equation}
where the number 62.000 has been obtained numerically
for a sufficiently large array ($1024 \times 1024$); at this size finite-size
effects have been confirmed to be negligible.
The critical value of $E_J/E_1$, below which the array is in
the insulating phase, is obtained from the condition
$E_{(ic)} - E_{(c)} = 0$: 
\begin{equation} \label{eq_3427}
\left(\frac{E_J}{E_1}\right)_c \approx 0.508 .
\end{equation}
Note that this value is somewhat larger than the critical value
$(E_J/E_1)_c \approx 0.23$ obtained in
Refs.~\onlinecite{bjkim,fazio}.  One possible explanation of the above
discrepancy is the failure  of the second-order perturbation expansion.
Namely, consideration of 
higher-order terms may eliminate the discrepancy. The other possibility
is that the CD type excitation does not give the true transition
point: See Fig.~\ref{fig_sch}~(b) for a schematic picture, where $K_1$ 
may correspond to the obtained transition point $(E_J/E_1)_c \approx 0.508$
and $K_2$ the true transition point. 
In experiment, the even larger critical value
$(E_J/E_1)_c \approx 0.6$ has been obtained~\onlinecite{zant},
via the curve-fit to the
square-root cusp form of the resistance~\cite{lobb}.
Recent quantum Monte Carlo simulations, on the other hand, have yielded the result 
$(E_J/E_1)_c = 0.36 \pm 0.04$. 

\subsection{Phase transitions for $\langle q_i \rangle \ne 0$} \label{subsec_pq}

In the NN charging limit, 
there exists an interesting duality between the charge and the 
vortex~\cite{fazio}, which concludes that
in the absence of the Josephson term, the charges form a superlattice
determined by the value of 
charge frustration $q_x$.
This is a counterpart of the vortex superlattice in the classical
array without the charging energy, formed in the presence of magnetic 
frustration~\cite{proceeding,straley}.
In the system considered here, on the other hand, the non-vanishing
self-capacitance gives the interaction range between charges
$\Lambda = \sqrt{C_1/C_0}$ in units of the lattice 
constant, thus making it finite~\cite{fazio}. 
Accordingly, we expect
that the formation of the $q\times q$ charge superlattice with 
$q \gg \Lambda$ is improbable in the system with the rational charge
frustration $q_x = p/q$. This implies that $\langle q_i \rangle$ is not
necessarily equal to $q_x$ in the 
array with non-vanishing $C_0$, which reflects that a short-ranged
potential in general makes the commensurate phase extend over a finite
range of the chemical potential. (Note that the charge frustration
in this work plays the role of the chemical potential.)

We now investigate the phase transitions for $\langle q_i \rangle =p/q$
$(\neq 0)$, using the canonical ensemble with the total number of 
charges fixed. To find the zero-temperature phase diagrams we need to know
the configurations of the commensurate states and of the lowest-lying
incommensurate states, which is a highly nontrivial problem in the 2D system 
considered here, especially for large $q$. 
For the 1D Josephson junction array with $\langle q_i \rangle = 1/q$,
the commensurate state, the first
excited state, and the resulting zero-temperature phase boundaries 
have been obtained via the perturbation expansion~\cite{odintsov}.  
We write the energy of the commensurate state with 
$\langle q_i \rangle = p/q \neq 0$ in the form
\begin{equation} \label{eq_l}
E_{(c)}(p/q)=E_{(c)}^{(0)}(p/q)+E_{(c)}^{(1)}(p/q)+E_{(c)}^{(2)}(p/q)+O(E_J^3), 
\end{equation}
where
the relations $\sum_i q_i = N\langle q_i \rangle = N p/q$ and 
$\sum_j {\widetilde C}_{ij}^{-1} = 1$ give 
$E_{(c)}^{(0)}(p/q) =
4E_0\sum q_i{\widetilde C}_{ij}^{-1}q_j - 8E_0q_xNp/q + 4E_0q_x^2N$.
In general, when $\langle q_i\rangle \neq 0$, the commensurate ground states
are degenerate, which makes it necessary to consider the matrix element:
\end{multicols}
\noindent\rule{0.5\textwidth}{0.1ex}\rule{0.1ex}{2ex}\hfill
\begin{eqnarray} \label{eq_h}
V_{lm} \equiv \langle {\bf q}^{(l)} | V | {\bf q}^{(m)} \rangle
= -\frac{E_J}{2}\sum_{\langle i,j\rangle}\left[\prod_{k\neq i,j}
  \delta\left(q_k^{(l)},q_k^{(m)}\right) \right] & & 
\left[ \delta\left(q_i^{(l)},q_i^{(m)} - 1\right) 
\delta\left(q_j^{(l)},q_j^{(m)} + 1\right) \right.   \nonumber \\
 & & + \left. \delta\left(q_i^{(l)},q_i^{(m)} + 1\right) 
\delta\left(q_j^{(l)},q_j^{(m)} - 1\right) \right] 
\end{eqnarray}
with $|{\bf q}^{(l)}\rangle$ being the $l$th
degenerate commensurate state. Note that the configuration of the
superlattice unit cell in a degenerate commensurate state is in general
different from that in a different (degenerate) state. Therefore, for
$l \neq m$, ${\bf q}^{(l)}$ differs
from ${\bf q}^{(m)}$ at the total of $O(N)$ sites for finite $q$, leading
to $V_{lm} = 0$. Further, it is obvious that $V_{ll} = 0$. It is
thus concluded that all the components of
the matrix $V$ vanish, which results in $E_{(c)}^{(1)}(p/q) = 0$.
The second-order contribution is given by
\begin{equation} \label{eq_i}
E_{(c)}^{(2)}(p/q) = {\sum_{\{{\bf q}'\}}}^\prime \frac
      {\left|\langle {\bf q}'|V|{\bf q}\rangle\right|^2}
      {E^{(0)}({\bf q}) - E^{(0)}({\bf q}')} ,
\end{equation}
where $E^{(0)}({\bf q})$ and $E^{(0)}({\bf q}')$ is the zeroth-order energy
for given charge configuration ${\bf q}$ and ${\bf q}'$, respectively, and
the summation ${\sum }^\prime$ is over the states ${\bf q}'$ making
the denominator nonzero.
The superscript $l$ on ${\bf q}$ has been dropped since all the degenerate
commensurate states give the same contribution to 
$E_{(c)}^{(2)}(p/q)$. Namely, the degeneracy is not removed in a finite
order since
${\bf q}^{(l)}$ and ${\bf q}^{(m)}$ cannot be connected by a finite number
of charge hoppings.
With Eq.~(\ref{eq_h}), Eq.~(\ref{eq_i}) reads 
\begin{equation}\label{eq_j}
E_{(c)}^{(2)}(p/q) = {\sum_{(i,j)}}^\prime 
\frac{E_J^2 / 4 } {E^{(0)}({\bf q}) - E^{(0)}({\bf q}_{(i,j)})} ,
\end{equation}
where ${\bf q}_{(i,j)} \equiv (q_1, q_2, \cdots ,$ $q_{i-1}, q_i{+}1, q_{i+1} 
\cdots,$ $q_{j-1}, q_j{-}1, q_{j+1} \cdots ,q_N)$ is obtained from
${\bf q}$ by a charge hopping between sites $i$ and $j$.
It is easy to compute the denominator in Eq.~(\ref{eq_j}):
\begin{eqnarray}
E^{(0)}({\bf q}) - E^{(0)}({\bf q}_{(i,j)}) &=& 4E_0\sum_{k,l}\left[
 (q_k -q_x){\widetilde C}_{kl}^{-1}(q_l -q_x) -
 (q_{(i,j),k} -q_x){\widetilde C}_{kl}^{-1}(q_{(i,j),l} -q_x)\right] \nonumber \\
&=& 8E_0\left[ \sum_l\left( {\widetilde C}_{jl}^{-1} -{\widetilde C}_{il}^{-1}
\right)q_l - \left( {\widetilde C}_{00}^{-1} - {\widetilde C}_{10}^{-1}\right)
\right] ,
\end{eqnarray}
which leads to the expression for the energy of 
the commensurate state
with $\langle q_i \rangle = p/q$:
\begin{eqnarray} \label{eq_o}
E_{(c)}(p/q) &\approx& 
          4E_0\sum_{i,j}q_i{\widetilde C}_{ij}^{-1}q_j - 8E_0q_xN\frac{p}{q}
                    + 4E_0q_x^2N \nonumber \\
& &+\frac{{E_J}^2}{32E_0}{\sum_{(i ,j)}}^\prime
   \frac{1}{\sum_l\left( {\widetilde C}_{jl}^{-1} -{\widetilde C}_{il}^{-1}
\right)q_l - \left( {\widetilde C}_{00}^{-1} - {\widetilde C}_{10}^{-1}\right)} .
\end{eqnarray}
Here $q_i$ is the charge at site $i$ in the commensurate state with $\langle
q_i \rangle = p/q$. Since the (commensurate) ground state configurations
are not known for general $p/q$, $q_i$ in this work is obtained via the
simulated annealing method.

To compute the energies of the incommensurate states with 
$\langle q_i\rangle = p/q \neq 0$,
we classify excitations according to the number $q_e$ of extra charges
(with the sign taken into account), and let $E_{(ic)}(p/q,q_e)$ denote
the energy of this excited state
which has the total number of charges $Np/q +q_e$. 
To the second order in $E_J$, it is straightforward to obtain the 
energy of the $q_e$-charge excited state:
\begin{eqnarray} \label{eq_Eic}
E_{(ic)}(p/q,q_e) &\approx& 
       4E_0\sum_{i,j}q_i^{(q_e)}{\widetilde C}_{ij}^{-1}q_j^{(q_e)} 
       - 8E_0q_x\left( N\frac{p}{q} + q_e \right) + 4E_0q_x^2N \nonumber \\
    & &+ \frac{{E_J}^2}{32E_0} {\sum_{(i,j)}}^\prime 
      \frac{1}{\sum_l\left( {\widetilde C}_{jl}^{-1} -{\widetilde C}_{il}^{-1}
       \right)q^{(q_e)}_l - \left( {\widetilde C}_{00}^{-1} - 
       {\widetilde C}_{10}^{-1}\right)} ,
\end{eqnarray}
\hfill\raisebox{-1.9ex}{\rule{0.1ex}{2ex}}\rule{0.5\textwidth}{0.1ex}
\begin{multicols}{2}
\noindent
where $q_i^{(q_e)}$ is the charge at site $i$ in the $q_e$-charge excited
state, and the first-order contribution $E_{(ic)}^{(1)}(p/q,q_e)$
has been observed to vanish in numerical calculations, regardless 
of the values of $p/q$  $(\ne 0)$ and $q_e$ considered in this work. Further,
in Eq.~(\ref{eq_Eic}), we have assumed that degeneracy is not removed in the
second order although there are some exceptions, e.g.,  for  
$p/q$=1/2 and $q_e = -1$ (see below).
To find the phase boundary,
we need to know the charge configuration which corresponds to the first 
excited state for given values of $C_1/C_0$ and $q_x$.  For this purpose,
we use the following numerical procedure:
\begin{enumerate}
\item{ Distribute $Np/q + q_e$ charges randomly.}
\item{Use the simulated annealing method to find the charge configuration
$q_i^{(q_e)}$, which minimizes the zeroth-order energy  
$4E_0\sum_{i,j}q_i^{(q_e)}{\widetilde C}_{ij}^{-1}q_j^{(q_e)} - 
  8E_0q_x (Np/q + q_e)$.}
\item{ Change the value of $q_e$, and repeat the steps 1 - 3.}
\item{ Among the obtained charge configurations with their zeroth-order
energies, find one corresponding to the lowest-lying
incommensurate state.}
\end{enumerate}
In the numerical simulations, we have used the conventional simulated
annealing method adopting the Metropolis algorithm for the arrays of
sizes $10 \times 10$ and $12 \times 12$. Starting from high
temperatures ($T > 1.0$),
we decrease the temperature slowly with the decrements
$\Delta T = 0.1$ (at high temperatures) and 0.002 (at
low temperatures), and perform 10000 MC steps per
site at each temperature step.
We have found that the charge configuration corresponding to the lowest
energy depends crucially on the value of $C_1/C_0$, and thus used three
different values: $C_1/C_0$ = 0.1, 1.0, and 5.0. 
Through 20 independent runs, we have obtained the charge
configurations corresponding to the lowest
energy, which have also been checked via the entropic annealing
algorithm~\cite{jlee}.
Although we have allowed double occupancy ($q_i = 2$) as well,
in the lowest-lying configurations
$q_i$ is always found to have the value either 0 or 1. 

Once the charge configuration $q_i^{(q_e)}$ which minimizes the zeroth-order
energy in Eq.~(\ref{eq_Eic}) is found, 
the phase boundary is determined from the condition
$E_{(c)}(p/q) = E_{(ic)}(p/q,q_e)$. This leads to the critical value
$q_x^c$ separating the superconducting phase from the insulating phase:
\end{multicols}
\noindent\rule{0.5\textwidth}{0.1ex}\rule{0.1ex}{2ex}\hfill
\begin{eqnarray} \label{eq_qxc}
q_x^c  \equiv \frac{1}{2q_e}& &\left[\sum_{i,j}{\widetilde C}_{ij}^{-1}(q_i^{(q_e)}q_j^{(q_e)}
 - q_i q_j)\right] \nonumber \\
   &+& \frac{(E_J/E_0)^2}{256q_e}\left[
      {\sum_{(i,j)}}^\prime 
      \frac{1}{\sum_l\left( {\widetilde C}_{jl}^{-1} -{\widetilde C}_{il}^{-1}
       \right)q^{(q_e)}_l - \left( {\widetilde C}_{00}^{-1} - 
       {\widetilde C}_{10}^{-1}\right)} \right.  \nonumber \\
    & & \left. -{\sum_{(i,j)}}^\prime 
      \frac{1}{\sum_l\left( {\widetilde C}_{jl}^{-1} -{\widetilde C}_{il}^{-1}
       \right)q_l - \left( {\widetilde C}_{00}^{-1} - 
       {\widetilde C}_{10}^{-1}\right)} \right] ,
\end{eqnarray} 
\hfill\raisebox{-1.9ex}{\rule{0.1ex}{2ex}}\rule{0.5\textwidth}{0.1ex}
\begin{multicols}{2}
\noindent
where $\{q_i \}$ again describes the charge configuration of the commensurate
state. When $q_e < 0$, the system, displaying superconductivity for
$q_x < q_x^c$, becomes insulating as $q_x$ is increased beyond $q_x^c$.
This behavior according to whether $q_x$ is larger or smaller than $q_x^c$
is reversed when $q_e > 0$.

\begin{figure}
\centerline{\epsfxsize=8.0cm \epsfbox{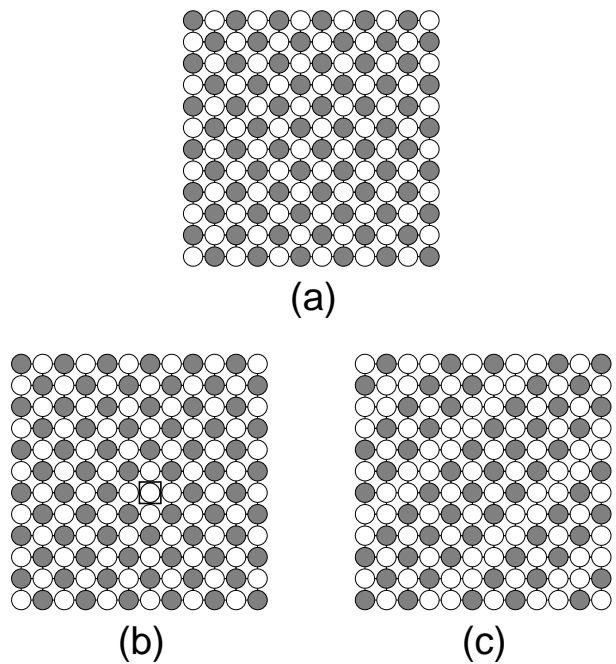}}
\vskip 0.5cm
\caption{(a) The commensurate charge configuration with 
$\langle q_i \rangle = 1/2$ at zero temperature, where the empty and
the filled circles denote the vacant ($q_i = 0$) and the occupied
(by single positive charges; $q_i = 1$) sites, respectively.
This commensurate state is found to be destroyed by a single negative charge 
excitation ($q_e = -1$) for $C_1/C_0 = 1.0$ and 5.0, as shown in (b), 
where the location of the point defect is indicated by a small square.
When $C_1/C_0$=0.1, on the other hand, the lowest excitation is found to have 
the configuration (c), which has  $q_e = -12$.}
\label{q2_conf}
\end{figure}

Table~\ref{tab_qe} shows the values of $q_e$ in the lowest excitations,
breaking the commensurability of the states with $\langle q_i \rangle$=
1/2, 1/3, 1/4, and 1/5, in the arrays of sizes $L=$ 10 and 12. 
Except for $\langle q_i \rangle=1/2$, there are two values
of $q_e$ in each case, one for the positive-charge excitation ($q_e>0$) and
the other for the negative-charge excitation ($q_e<0$). Since the value
of the charge
frustration has been restricted in the range $0 \le q_x < 1/2$, we need to consider only the
negative-charge excitations for $\langle q_i \rangle = 1/2$.
It displays that as $C_1$ is increased,
the absolute value of $q_e$ decreases, manifesting that the excitations become more 
point-like. As an example, we present the charge configurations for $\langle
q_i \rangle = 1/2$ in Fig.~\ref{q2_conf}, where (a), (b), and (c)
correspond to the commensurate state, the excited state with $q_e = -1$,
and the excited state with $q_e = -12$, respectively, in a
$12\times 12$ array. It is of interest to note that the configuration 
(c), which has 60 charges,
is precisely the same as the ground state configuration of vortices in the 
presence of the magnetic frustration 5/12~\cite{straley}.
When $q_e = -1$ is the lowest excitation, there exists $N/2$-fold degeneracy
since the vacant site, represented by a small square in (b),
can be located on any of the total $N/2$ sites. In this case,
the degenerate states are coupled in $O(E_J^2)$, and 
we should use the second-order degenerate perturbation theory to calculate
the energy.
It is again straightforward to 
diagonalize the second-order matrix by means of the Fourier 
transform, which yields 
\end{multicols}
\noindent\rule{0.5\textwidth}{0.1ex}\rule{0.1ex}{2ex}\hfill
\begin{eqnarray} \label{eq_deg}
E_{(ic)}(1/2,-1) & \approx &
       4E_0\sum_{i,j}q_i^{(-1)}{\widetilde C}_{ij}^{-1}q_j^{(-1)}
       - 8E_0q_x\left( \frac{1}{2}N - 1 \right) + 4E_0q_x^2N \nonumber \\
    & &+ \frac{{E_J}^2}{32E_0} \left[ {\sum_{(i,j)}}^\prime
      \frac{1}{\sum_l\left( {\widetilde C}_{jl}^{-1} -{\widetilde C}_{il}^{-1}
       \right)q^{(-1)}_l - {\widetilde C}_{00}^{-1} +
       {\widetilde C}_{10}^{-1}} \right. \nonumber \\
    & &  + \frac{12}
       {\sum_l\left( {\widetilde C}_{k+{\hat {\bf x}},l}^{-1} -
            {\widetilde C}_{kl}^{-1}
       \right)q^{(-1)}_l - {\widetilde C}_{00}^{-1} +
       {\widetilde C}_{10}^{-1}}  \nonumber \\
    & & +  \frac{8}
       {\sum_l\left( {\widetilde C}_{k+{\hat {\bf x}}+{\hat {\bf y}},l}^{-1} -
            {\widetilde C}_{k + {\hat {\bf x}},l}^{-1} 
       \right)q^{(-1)}_l -  {\widetilde C}_{00}^{-1} +
       {\widetilde C}_{10}^{-1}}  \nonumber \\
    & &  + \left. \frac{4} 
       {\sum_l\left( {\widetilde C}_{k+2{\hat {\bf x}},l}^{-1} -
            {\widetilde C}_{k+{\hat {\bf x}},l}^{-1} 
       \right)q^{(-1)}_l - {\widetilde C}_{00}^{-1} +
       {\widetilde C}_{10}^{-1}} \right] ,
\end{eqnarray}
\hfill\raisebox{-1.9ex}{\rule{0.1ex}{2ex}}\rule{0.5\textwidth}{0.1ex}
\begin{multicols}{2}
\noindent
in place of Eq.~(\ref{eq_Eic}). 
Here $k$ denotes the position of the vacant site.

\begin{figure}
\centerline{\epsfxsize=12.0cm \epsfbox{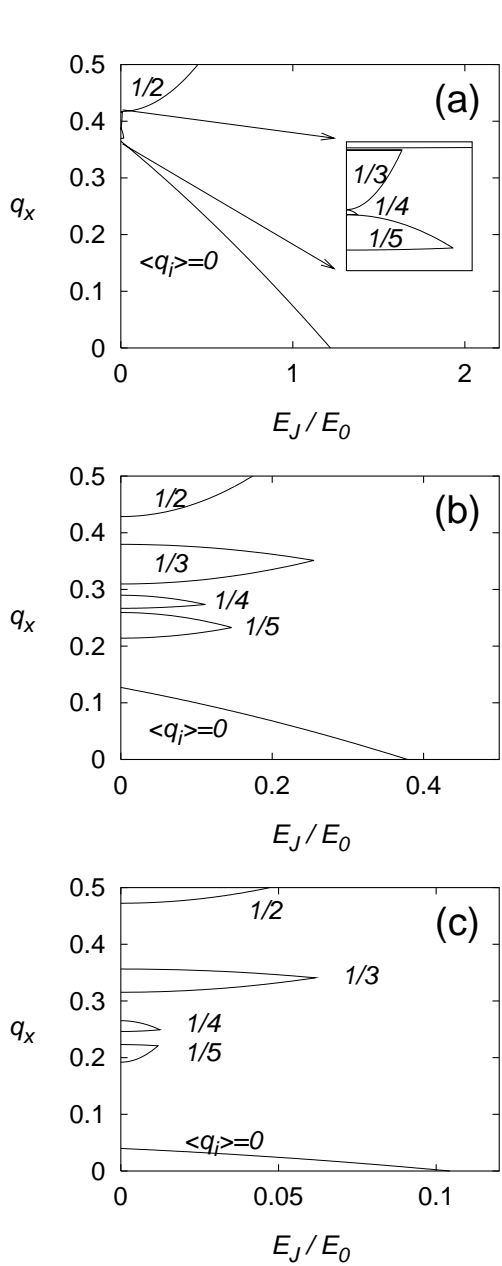}}
\caption{The zero-temperature phase boundaries for $C_1/C_0$ = 
(a) 0.1, (b) 1.0, and (c) 5.0,  obtained via the second-order perturbation
expansion, with the help of the simulated annealing method to find the
charge configuration. Various lobe-like structures, inside of  which the array
is in the insulating phase, are found in both the $10\times 10$ and the 
$12\times 12$ array.
As we increase the array size, additional insulating lobes are expected to be
observed. As $C_1/C_0$ is increased, the insulating lobes become narrower, 
with the central position approaching $\langle q_i \rangle = q_x$. In the inset of 
(a), the insulating lobes with $\langle q_i \rangle =$ 1/3, 1/4, and 1/5 
are shown. 
}
\label{fig_zeroT}
\end{figure}

Figure~\ref{fig_zeroT} shows the resulting phase diagrams for $C_1/C_0$=
(a) 0.1, (b) 1.0, and (c) 5.0. The various lobe-like structures, in which
the array is in insulating phase, are obtained. Since the screening length
$\Lambda = \sqrt{C_1/C_0}$, beyond which the magnitude of the
interaction between charges decreases exponentially, is much smaller
than the system size, the finite-size effects on the 
phase boundaries are presumably negligible. We have checked the phase
boundary for $\langle q_i \rangle = 1/2$, and indeed found that it
hardly changes with the system size
considered here (see below). Accordingly, the phase boundaries in
Fig.~\ref{fig_zeroT} are expected to be qualitatively correct even
in the thermodynamic limit. As we increase the array size, more
insulating-lobes should be observed, 
since the array with $E_J = 0$ should be insulating, regardless of the
value of $q_x$. At this stage, however, it is still unclear
whether the insulating lobe exists at every rational value of $q_x$.
In the NN charging limit, the charging-energy term takes the form
$ H_0  = 4E_1\sum_{i,j} (q_i - q_x){\bar C}_{ij}^{-1}(q_j-q_x)$
with ${\bar C}_{ij}^{-1}$ given by Eq.~(\ref{eq_Cbar}).
Here the divergence of ${\bar C}_{00}^{-1}$ leads to 
$\langle q_i \rangle = q_x$, which is simply
the counterpart of the vortex-neutrality condition in the classical
array. Indeed Fig.~\ref{fig_zeroT} shows that
the position of the insulating region  
approaches $\langle q_i \rangle = q_x$, as $C_1/C_0$ is increased.

To study the effects of the junction capacitance versus the self-capacitance
in detail, we now concentrate on the
$\langle q_i \rangle = 1/2$ insulating phase, whose commensurate charge
configuration displays the $2\times 2$ superlattice structure.
[See Fig.~\ref{q2_conf}~(a).] Similarly to the vortex superlattice
in the fully-frustrated classical array~\cite{mychoi85}, the
commensurate state has the two-fold degeneracy.
If we remove a single positive charge from the system, commensurability is
destroyed near the resulting point defect. The charge configuration
of the corresponding lowest-lying incommensurate state
has been shown in Fig.~\ref{q2_conf}~(b), which describes the first excited
state for $C_1/C_0$ = 1.0 or 5.0. When $C_1/C_0$ =
0.1, in contrast, the first excited state has the configuration
shown in Fig.~\ref{q2_conf}~(c), 
with the number of extra charge $q_e = -12$.
This observation that the nature of the lowest excited state changes with $C_1/C_0$ 
has an important implication with regard to the supersolid phase, 
where a commensurate 
charge-density wave is believed to coexist with 
superfluidity~\cite{meisel,otterlo}: 
The charge configuration in Fig.~\ref{q2_conf}~(b) regarded
as a snapshot of the charge configuration in the supersolid phase
demonstrates that the 
underlying global commensurability still holds despite of the 
local point defect. Furthermore, since the point defect
can hop to any site without energy cost, it is 
delocalized in the energy eigenstate, leading to superfluidity.
Accordingly, we conclude that the existence of the supersolid phase 
depends on the value of $C_1/C_0$. In particular, when $C_1/C_0$ is
sufficiently small, the supersolid phase does not exist. 
It is of interest to compare this argument pointing out the
importance of the vacant defect in supersolidity with that of
Ref.~\onlinecite{otterlo}, emphasizing the role of the double occupancy.
In the numerical investigations to find the lowest-energy charge
configurations, we have found that all the charges have the
value either 0 or 1 at zero temperature, and that those configurations
with  $q_i = -1$ or 2 at some sites have much higher energies.
The $\langle q_i \rangle = 1/2$ phase boundaries in Fig.~\ref{fig_zeroT}~(b) 
and (c) may be considered to divide the insulating commensurate phase and 
the supersolid phase. As we go further into the supersolid
phase starting from the phase boundary, the more point defects are generated;
the commensurability should be completely destroyed when we cross another
phase boundary separating the supersolid phase from the superfluid phase,
although within the formalism adopted here, we are unable to find this
new boundary. To obtain the phase diagram for $\langle q_i \rangle = 1/2$,
we have considered $L\times L$ arrays up to $L=24$, 
and found that the finite-size effects are negligible for $C_1/C_0
\lesssim 10$, which corresponds to the interaction range $\Lambda \lesssim 
\sqrt{10} \ll L$,  and that the qualitative features of the phase diagram
do not change even when $C_1/C_0 \gtrsim 10$.

\begin{figure}
\centerline{\epsfxsize=8.0cm \epsfbox{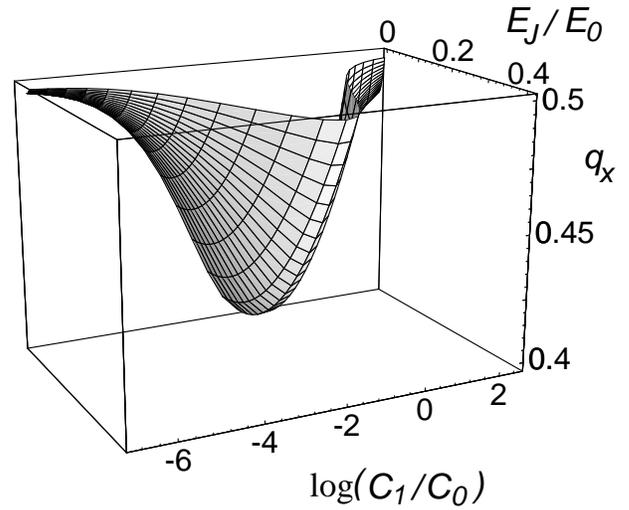}}
\vskip 0.5cm
\caption{
Zero-temperature phase diagram for $\langle q_i \rangle = 1/2$,
where only single negative charge excitation ($q_e = -1$) has been considered.
In the region above the surface, charges form a $2\times 2$ superlattice,
as shown in Fig.~\ref{q2_conf}~(a), leading to 
the insulating phase. As $C_1/C_0$ is increased, the insulating region expands 
at first ( for  $C_1/C_0 \protect\lesssim 0.1$) and then shrinks.}
\label{q2_bound}
\end{figure}

We have further found that the difference between the phase boundary
obtained from the excitation $q_e = -1$ and that from $q_e = -12$
is insignificant in the $12\times 12$ array with $C_1/C_0 = 0.1$. 
Therefore, for simplicity, we consider only the excitation $q_e = -1$ and compute the 
phase boundary from Eqs.~(\ref{eq_o}) and (\ref{eq_deg}).
The resulting phase diagram obtained for the array of size $L=24$ 
is shown in Fig.~\ref{q2_bound}. The insulating region is shown to expand 
with $C_1/C_0$ when $C_1/C_0 \protect\lesssim 0.1$ and then 
to shrink as $C_1/C_0$ is increased further. This is to be compared with the
case $\langle q_i \rangle = 0$ (shown in Fig.~\ref{q0_bound}), where 
superconductivity is enhanced monotonically with $C_1/C_0$, even for
small $C_1/C_0$. 

\section{Finite-temperature phase diagrams} \label{sec_finite} 
In this section, we investigate the phase transitions of the superconducting
arrays at finite temperatures, where thermal fluctuations as well as quantum
fluctuations suppress the ordering of phases. 
While at high temperatures the system should be in the
normal (disordered) state due to large thermal fluctuations,
strong charging effects at low
temperatures produce large quantum fluctuations, tending to destroy
superconductivity of the system.  We thus expect rich behaviors
resulting from the interplay of thermal and quantum fluctuations, which
may be controlled by the temperature and charging energy. 
We first study the system in the absence of the Josephson coupling via MC 
simulations,
and then investigate the phase transitions in the presence of the Josephson
coupling via a variational method together with MC simulations. 

\subsection{Superconducting array without Josephson coupling} 
\label{subsec_mc}
In the absence of the Josephson coupling ($E_J=0$), the Hamiltonian
is diagonal in the charge representation and written in the form
\begin{equation}
H = 4E_0\sum_{i,j}q_i{\widetilde C}_{ij}^{-1}q_j - 8E_0q_x\sum_i q_i ,
\end{equation}
where $q_i$ is the eigenvalue of the charge operator at site $i$.
We perform MC simulations on the above Hamiltonian to investigate how the
insulating phase corresponding to the regions inside the lobes in
Fig.~\ref{fig_zeroT} changes with the temperature.
Using the conventional MC method with the Metropolis
algorithm, we start from a sufficiently high temperature ($T=1.0$), 
and decrease the temperature slowly with the decrements
$\Delta T = 0.1$ (at high temperatures) and 0.002 (at low temperatures), 
During the simulations,
we allow creation and annihilation of a charge at a site, as well 
as hopping to other sites.
At each temperature step, we compute the thermal average 
$\langle q_i \rangle_T$  using 10000 MC steps per site,
after 1000 MC steps of equilibration.

\begin{figure}
\centerline{\epsfxsize=7.5cm \epsfbox{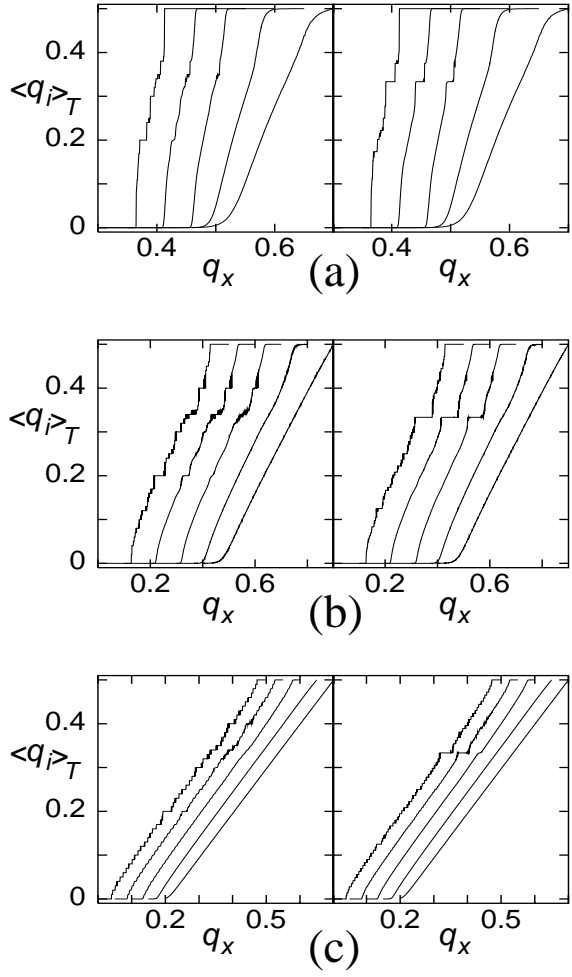}}
\vskip 0.5cm
\caption{
Monte Carlo results of $\langle q_i \rangle_T$ 
as functions of $q_x$ for arrays of sizes $L=10$ (left) and 12 (right) with 
$E_J = 0$ and $C_1/C_0 = $ (a) 0.1,
(b) 1.0, and (c) 5.0. From the left to the right, the curves correspond to 
temperatures $T=0.0, 0.006, 0.01, 0.05$, and 0.1 in (a), 
$T=0.0, 0.01, 0.014, 0.04$, and 0.1 in (b), and
$T=0.0, 0.004, 0.008, 0.02$, and 0.05 in (c), respectively.
Here the temperature has been measured in units of $E_0/k_B$, and all 
the curves have been shifted in the horizontal direction for clarity.
More steps are expected to be observed in systems of larger sizes. }
\label{fig_mc}
\end{figure}

Figure~\ref{fig_mc} displays the results for arrays of sizes
$L=10$ (left) and 12 (right), with $C_1/C_0 = $ (a) 0.1,
(b) 1.0, and (c) 5.0. At zero temperature, the curves representing
$\langle q_i \rangle$ versus $q_x$ exhibit steps at rational values
giving the (commensurate) insulating phase. Such steps structures,
which correspond to the lobe structures in Fig.~\ref{fig_zeroT}, are
common in systems displaying commensurate-incommensurate transitions,
and signal the commensurate locking~\cite{topol}.
It is shown that, as expected, the width of a step decreases with
$C_1/C_0$, and the curve approaches the straight line $\langle q_i \rangle
= q_x$ in the limit $C_1/C_0 \rightarrow \infty$, manifesting the
suppression of the insulating phase. Comparing the results for the
$L=10$ array with those for the $L=12$ array, we also expect more
rational steps in larger arrays, which is apparently suggestive of
the devil's staircase structure~\cite{topol} in the thermodynamic
limit. For conclusive results, however, more detailed study is needed.
Table~\ref{tab_mc} presents the widths of the  
rational steps of $\langle q_i \rangle = 0,$ 1/3, and 1/2 in the
$L=12$ array at zero temperature. One can see here that 
the values in Table~\ref{tab_mc} are in good agreement with 
the $E_J=0$ results in Fig.~\ref{fig_zeroT}, which supports the validity
of the approach in Sec.~\ref{sec_zero}.

Figure~\ref{fig_mc} also shows that steps tend to disappear at finite
temperatures, reflecting that thermal fluctuations suppress the charge
ordering due to quantum coherence as well as the phase ordering.
Accordingly, the insulating commensurate phase in general turns into the
normal (disordered) phase as the temperature is increased from zero.
Note, however, that some steps survive weak thermal fluctuations.
Namely, whereas steps at higher-order rationals easily disappear, low-order
steps such as $\langle q_i \rangle = 0$, 2, and 1/3 persist at low but
nonzero temperatures.
These features presumably can also be observed in the systems with nonzero
Josephson couplings.  Thus as we increase the temperature from zero,
higher-order insulating lobes in Fig.~\ref{fig_zeroT} disappear, but
some low-order lobes are expected to remain at low temperatures.

\subsection{Superconducting array with Josephson coupling} \label{subsec_var}

In the presence of the Josephson coupling, both the charge and the phase
variables should be treated quantum-mechanically, which prohibits exact
analytical treatment. Here, we use the Giachetti-Tognetti-Feynman-Kleinert (GTFK) 
variational method~\cite{GTFK} to evaluate the path integral and to obtain 
the effective classical Hamiltonian, which is convenient for studying
the phase transitions at finite temperatures.  
The GTFK method, which remains reliable in known 
cases even at zero temperature, has been successfully applied to the
superconducting arrays with continuous charge states in the presence of Ohmic 
dissipation~\cite{bjkim,skim}.
We begin with the Hamiltonian given by Eq.~(\ref{eqH2}) and write the partition 
function of the system in terms of the path integral~\cite{feynman}:
\begin{equation}\label{EqZ}
Z = \prod_i \int d\phi_i(0) \int{\cal D} \phi_i(\tau)
    \exp \left\{ -S_E[\phi_i(\tau)] \right\}
\end{equation}
with the Euclidean action
\begin{equation}\label{Eq:S_E}
S_E = \int_0^\beta d\tau L_E(\tau) ,
\end{equation}
where $\beta \equiv 1/k_BT$ is the inverse temperature, and the Planck constant
has been set equal to unity ($\hbar \equiv 1$). The Euclidean 
Lagrangian $L_E$ can be obtained from the Hamiltonian via the 
Wick rotation $ t \longrightarrow -i\tau $ and the Legendre 
transformation. We use the representation 
$q_i = -i \partial / \partial \phi_i$ obeying 
the commutation relation $[\phi_i,q_i] = i$, and get
\end{multicols}
\noindent\rule{0.5\textwidth}{0.1ex}\rule{0.1ex}{2ex}\hfill
\begin{equation}
L_E = \frac{1}{16E_0} \sum_{i, j}\dot{\phi_i} {\widetilde C}_{ij} 
      \dot{\phi_j} - iq_x \sum_i \dot{\phi_i}
       - E_J \sum_{\langle i, j \rangle} \cos(\phi_i - \phi_j)
\end{equation}
with $ \dot{\phi_i} \equiv \partial \phi_i / \partial \tau $,
where the charge frustration enters through a purely imaginary term.

In the presence of Ohmic dissipation, the charge at each site takes
continuous values, and only the paths satisfying $\phi_i(\beta) = \phi_i(0)$
contribute to the partition function, with the dissipative action term 
added~\cite{bjkim,schon}. In this case, the charge frustration
does not appear in the Euclidean action since $\int_0^\beta d\tau 
\dot{\phi_i} = 0$, and the resulting phase diagram does not depend on $q_x$.
If there is no Ohmic dissipation present, in contrast, only discrete charge states 
are allowed, and all the paths satisfying $ \phi_i(\beta) = \phi_i(0) + 
2\pi n_i $ with integer $n_i$ contribute to the path integral in 
Eq.~(\ref{EqZ})~\cite{schon}.  Therefore the path integral should include
the summation over the ``winding numbers'' $\{n_i\}$, yielding interesting 
behaviors associated with the charge frustration~\cite{bruder}. 
Accordingly, the allowed charge states,  i.e., continuous or discrete, 
are crucial in the resulting phase diagrams. 

From the boundary condition $ \phi_i(\beta) = \phi_i(0) + 2\pi n_i $,
we decompose $\phi_i$ into the periodic variable $\theta_i$ satisfying  
$ \theta_i(\beta) = \theta_i(0)$ and $n_i$, according to  
$\phi_i(\tau) \equiv \theta_i(\tau) + (2\pi\tau/\beta) n_i$, 
and obtain the Euclidean action in the 
form
\begin{equation}
S_E  =  \frac{\pi^2}{4\beta E_0} \sum_{i,j} n_i {\widetilde C}_{ij} n_j
       - i 2 \pi q_x \sum_i n_i + S_{ph} , 
\end{equation}
where
$$
S_{ph}   \equiv  \label{Eq:S}
    \int_0^\beta d\tau \left\{\frac{1}{16E_0} \sum_{i,j} \dot{\theta_i} 
     {\widetilde C}_{ij} \dot{\theta_j} - E_J \sum_{\langle i,j \rangle}
    \cos \left[ \theta_i - \theta_j + 
        \frac{2 \pi \tau}{\beta}(n_i - n_j)\right] \right\} .
$$
Following the GTFK method, it is 
straightforward to obtain 
the effective classical Hamiltonian 
\begin{equation}\label{eqHcl}
-\beta H_{cl} = - \frac{\pi^2}{4\beta E_0} \sum_{i,j} n_i 
               {\widetilde C}_{ij} n_j + i 2 \pi q_x \sum_i n_i +
               g \beta E_J \sum_{\langle i,j \rangle}
              \cos(\theta_i - \theta_j) \, \delta_{n_i,n_j},
\end{equation}
where $g$ is determined by
\begin{eqnarray}\label{Eq:g}
g &=& g_0\; (1 - \log g_0) \nonumber \\
-\log g_0 &=& \frac{\beta E_0}{\pi^2} \sum_{n=1}^\infty
   \left\{\left[\left(1 + \frac{2C_1}{C_0}\right)n^2 + 
           \frac{4g_0\beta^2 E_J E_0 }
          {\pi^2}\right\}^{-1} \right. \nonumber\\
& & \quad \quad \quad \quad \left.   +
\left\{\left(\frac{1}{3}+ \frac{2C_1}{C_0}\right)n^2 
      + \frac{4g_0\beta^2 E_J E_0 }{\pi^2}\right]^{-1}
\right\} \nonumber\\
&=&\label{Eq:g0} \frac{1}{8 g_0 \beta E_J }
(x_1 \coth x_1 + x_2 \coth x_2 - 2)
\end{eqnarray}
with $x_1 \equiv \sqrt{4g_0 \beta^2 E_J E_0  /(1 + 2C_1/C_0)}$ and
$x_2 \equiv \sqrt{4g_0 \beta^2 E_J E_0  /(1/3 + 2C_1/C_0)}$. 
Here we have used identity 
$$
\sum_{n=1}^\infty \frac{1}{n^2 + (x/\pi)^2}
    = \frac{\pi^2}{2 x^2}(x\coth x - 1).
$$
When Eq.~(\ref{Eq:g0}) has more than one zeros, we should choose the largest 
one according to the extremal principle.
The detailed procedure is entirely similar to that in Ref.~\onlinecite{bjkim},
and will not be repeated here. 
As a result, the quantum fluctuations renormalize the Josephson coupling 
$E_J$ to $gE_J$.
Furthermore, in contrast to the continuous charge case~\cite{bjkim},
the charge frustration appears explicitly in the Hamiltonian,
and is expected to play an important role
in the phase transition of the system.

Unfortunately,  
the Kronecker delta $\delta_{n_i,n_j}$ between winding numbers of 
nearest neighboring pairs in Eq.~(\ref{eqHcl}) makes further analysis 
very difficult.
For simplicity,  we thus replace it by its self-consistent average $\delta$
and obtain the approximate Hamiltonian
\begin{equation}\label{eqHcl'}
-\beta H_{cl}^{\prime} = - \frac{\pi^2}{4\beta E_0} \sum_{i,j} n_i
               {\widetilde C}_{ij} n_j + i 2 \pi q_x \sum_i n_i +
    g \delta \beta E_J \sum_{\langle i,j \rangle} \cos(\theta_i - \theta_j) ,
\end{equation}
where
$$
\delta \equiv \langle\delta_{n_i,n_j}\rangle 
    \equiv \frac {\hbox{Tr } \delta_{n_i,n_j} e^{-\beta H_{cl}^{\prime}}}
        {\hbox{Tr } e^{-\beta H_{cl}^{\prime}}}
$$
is the ensemble average with respect to the approximate Hamiltonian 
$H_{cl}^{\prime}$.
The Poisson summation formula then allows us to write
\begin{equation}\label{eqde}
\delta = \frac {\int_0^1 dx \sum_{\{m_i\}}
    \exp \left[ - 4\beta E_0
        \sum_{i,j}\Biglb( m_i - q_x -q_i^{\prime}(x)\Bigrb)
    {\widetilde C}_{ij}^{-1}\Biglb( m_j - q_x - q_j^{\prime}(x)\Bigrb) \right] }
    {\sum_{\{m_i\}}
    \exp \left[ - 4\beta E_0
    \sum_{i,j} (m_i - q_x) {\widetilde C}_{ij}^{-1} (m_j - q_x) \right]}
\end{equation}
\hfill\raisebox{-1.9ex}{\rule{0.1ex}{2ex}}\rule{0.5\textwidth}{0.1ex}
\begin{multicols}{2}
\noindent
with $q_i^{\prime}(x) \equiv x (\delta_{i,1} - \delta_{i,2}) $, 
which manifests the absence of dependence on $E_J$.
Once $\delta$ is determined from Eq.~(\ref{eqde}), the partition
function of the system obtains the form 
\begin{equation}
Z = Z_{ch} Z_{ph}, 
\end{equation}
where the charge-part
$$
Z_{ch} = \sum_{ \{q_i\} } \exp\left[ -4\beta E_0\sum_{i,j} (q_i-q_x)
      {\widetilde C}_{ij}^{-1} (q_j - q_x) \right]
$$
has been obtained from the Poisson summation formula and the phase-part
is given by
$$
Z_{ph} = \int \prod_i d\theta_i \exp\left[ -g\delta\beta E_J
     \sum_{\langle i,j\rangle}\cos(\theta_i - \theta_j)\right].
$$
Unless the ratio $C_1/C_0$ is infinite, the interaction
$\widetilde{C}_{ij}^{-1}$ between charges is short-ranged and does not
have any singularity. It is then reasonable to assume that the criticality
of the system is governed by the phase-part, which implies that
the system exhibits a vortex-unbinding BKT transition at the critical
temperature given by~\cite{soh}
\begin{equation}\label{eqcritical}
g \delta \beta E_J \approx 1/0.9 .
\end{equation}

\begin{figure}
\centerline{\epsfxsize=10.0cm \epsfbox{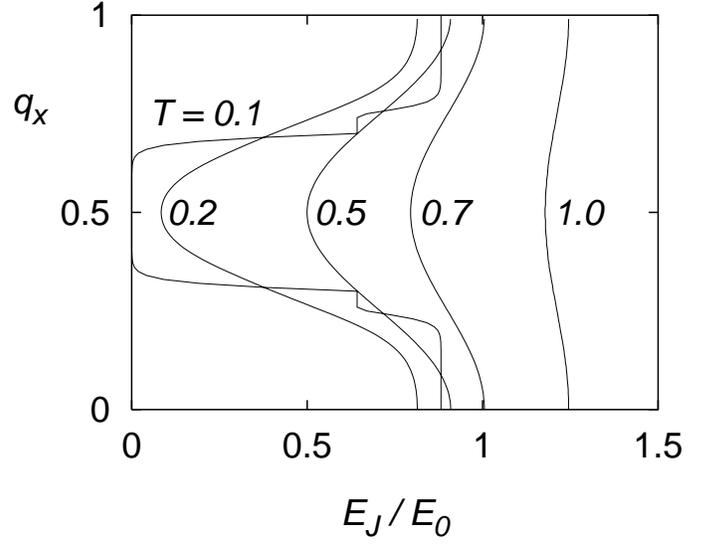}}
\vskip 0.5cm
\caption{Phase diagrams in the self-charging limit ($C_1= 0$) at temperatures
$T=0.1$, 0.2, 0.5, 0.7, and 1.0. Superconducting region (the right-hand
side of each curve) is shown to expand 
as $q_x$ is increased. Reentrant behavior is observed at $q_x=0$.
}
\label{fig_selfbd}
\end{figure}

We first consider the self-charging limit ($C_1 = 0$). 
In this limit, $\delta$ can  easily be evaluated since the terms 
in Eq.~(\ref{eqde}) factorize.
The resulting phase diagrams determined from Eq.~(\ref{eqcritical}) 
at temperatures $T  = 0.1, 0.2, 0.5, 0.7$, and $1.0$ (in units
of $E_0/k_B$) are shown in Fig.~\ref{fig_selfbd}.
The region in the right-hand side of each line corresponds to the 
superconducting phase whereas the left region represents the 
normal (disordered) phase.
In this phase diagram, we observe that the superconducting region expands 
as $q_x$ is increased at all temperatures.
(Recall that the symmetry of the system allows to restrict within
the range [0,1/2].)
Therefore it is concluded that the charge frustration in general enhances 
superconductivity. The qualitative features are largely  similar to 
those of the results from the coarse-graining approach~\cite{bruder} at temperatures
$T > 0.2$.  Unlike the latter, however, Fig.~\ref{fig_selfbd} also
shows  reentrant behavior at low temperatures, for small $q_x$.
At zero temperature, unfortunately, we obtain the 
unphysical result that the system is in the superconducting state 
for $1/3 \le q_x \le 2/3$, even in the 
limit of the vanishing Josephson coupling.
This is presumably an artifact of the self-consistent approximation (SCA)
replacing of the Kronecker delta by its average, with correlations
neglected. 

\begin{figure}
\centerline{\epsfxsize=10.0cm \epsfbox{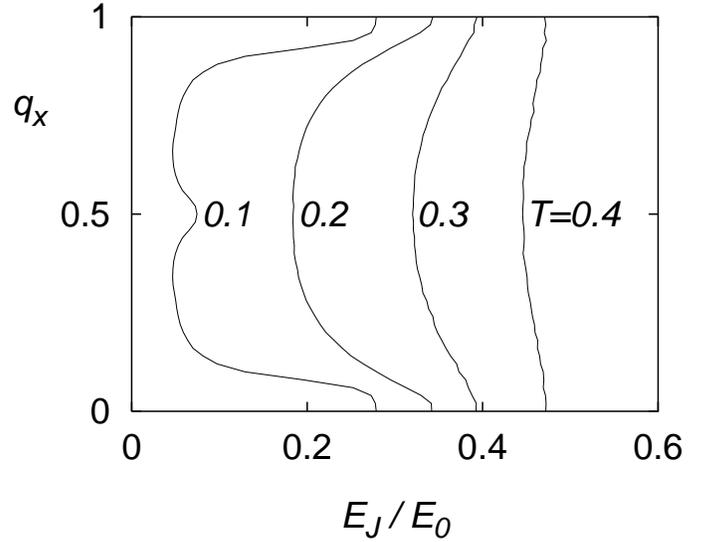}}
\vskip 0.5cm
\caption{Phase diagrams for $C_1/C_0 = 1.0$ at temperatures $T = 0.1$, 0.2,
0.3, and 0.4.  At $T=0.1$, there exists a lobe-like structure 
near $q_x=0.5$.}
\label{fig_1.0bd}
\end{figure}

When $C_1/C_0 \neq 0$, on the other hand, 
the terms in Eq.~(\ref{eqde}) do not factorize, 
and the average $\delta$ cannot be evaluated analytically,
which makes it inevitable to resort to a numerical method.
At extremely low temperatures, $m_i$ in Eq.~(\ref{eqde}) can have the value  
either 0 or 1, which allows us to calculate
$\delta$  by direct summation of $2^N$ terms for small $L$.
At high temperatures, other values of $m_i$ are allowed; this makes
the direct summation impractical and we use MC simulations.
Figure~\ref{fig_1.0bd} shows the resulting phase diagram obtained 
for a $4\times 4$ array with $C_1/C_0 = 1.0$
at temperature $T=0.1$, 0.2, 0.3, and 0.4.
At $T = 0.1$, $\delta$ has been calculated by means of both the direct summation 
with $m_i = 0$ and $1$ and MC simulations, which give results in
good agreement with each other. At higher temperatures, $\delta$ has been obtained
via MC simulations, yielding the enhancement of superconductivity by charge 
frustration in a similar manner to that in the self-charging limit. 
At low temperatures ($T=0.1$), on the other hand, there exists a 
small lobe near $q_x = 1/2$, which indicates the expansion of 
the insulating region due to the commensurability effects on 
the ordering of charges.

To check the validity of the SCA adopted here, replacement of
$\delta_{n_i, n_j}$ to its self-consistent value $\delta$, we have
performed MC simulations with $q_x = 0$, at which the Hamiltonian in 
Eq.~(\ref{eqHcl}) has only real terms:
\end{multicols}
\noindent\rule{0.5\textwidth}{0.1ex}\rule{0.1ex}{2ex}\hfill
\begin{equation} \label{eq_Hcl}
-\beta H_{cl} = - \frac{\pi^2}{4\beta E_0} \sum_{i,j} n_i 
               {\widetilde C}_{ij} n_j  +
               g \beta E_J \sum_{\langle i,j \rangle}
              \cos(\theta_i - \theta_j) \, \delta_{n_i,n_j} .
\end{equation}
We have considered a $16\times 16$ array with various values of $E_J/E_0$,
and measured the helicity modulus $\Upsilon$ with 100000 MC steps:
\begin{eqnarray}
    \Upsilon
        & \equiv & \frac{1}{N} \left[ 
            \left< \sum_{\langle i,j \rangle} x_{ij}^2
            \cos(\theta_i - \theta_j) \right> 
        - \beta E_J \left< \left( \sum_{\langle i,j \rangle} x_{ij}
                \sin(\theta_i - \theta_j) \right)^2 \right> \right. \nonumber\\
    & &     + \beta E_J \left.
            \left<  \sum_{\langle i,j \rangle} x_{ij}
                \sin(\theta_i - \theta_j) \right>^2
            \right],
    \end{eqnarray}
\hfill\raisebox{-1.9ex}{\rule{0.1ex}{2ex}}\rule{0.5\textwidth}{0.1ex}
\begin{multicols}{2}
\narrowtext
\noindent
where $x_{ij} \equiv x_j - x_i$
with $x_i$ denoting the $x$ coordinate of the $i$th grain in units of the lattice 
constant.
The universal jump condition of the helicity modulus~\cite{nelson}
\begin{equation}
    \Upsilon = 2 / \pi \beta E_J = 2\, T E_0 / \pi E_J 
\end{equation}
then determines the critical coupling $(E_J/E_0)_c$, the values of which are
displayed in Table~\ref{tab_j}, for $C_1/C_0=0$ and 1 at various 
temperatures. 
Table ~\ref{tab_j} also displays the results from the SCA replacing 
the Kronecker delta by its average, which has yielded Figs.~\ref{fig_selfbd}
and \ref{fig_1.0bd}: The two results indeed exhibit behaviors qualitatively
in accord with each other. In particular, for $C_1/C_0 = 0$, both
results give the value of $(E_J/E_0)_c$ larger at $T=0.1$ than at $T=0.2$, 
strongly suggesting the reentrant transition.

\begin{figure}
\centerline{\epsfxsize=10.0cm \epsfbox{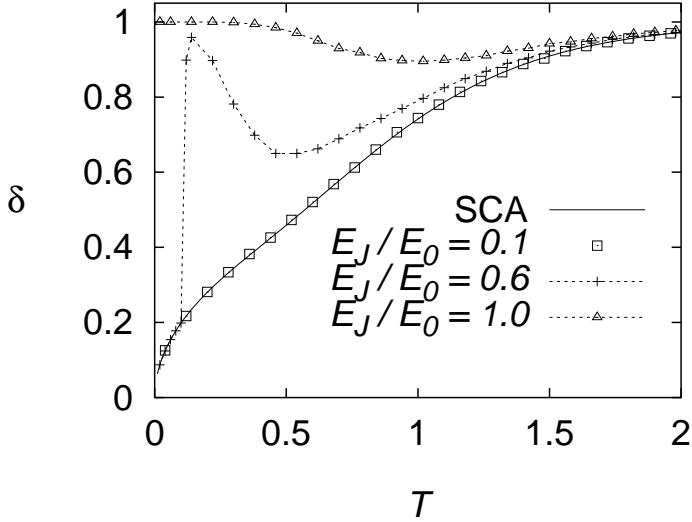}}
\vskip 0.5cm
\caption{
Monte Carlo results of $\delta \equiv\langle \delta_{n_i,n_j}\rangle$ 
for $E_J/E_0 = 0.1$, 0.6, and 1.0, at $q_x=0$ in the
self-charging limit ($C_1=0$). Results from 
the self-consistent approximation (SCA), which are independent of $E_J/E_0$,  
are also shown by the solid line. 
When the temperature is sufficiently high, the SCA is shown to give
reliable results; at low temperatures, SCA results are
in agreement with the MC results only for sufficiently small
$E_J/E_0$. Dotted lines are merely guides to the eye.
}
\label{fig_delta}
\end{figure}

From a quantitative viewpoint, on the other hand, the agreement between
the two results is not satisfactory, especially at low temperatures,
which seems to indicate that the SCA fails at low temperatures. 
To examine this, we have also measured in the MC simulations 
the average value of $\delta_{n_i,n_j}$ for $C_1 = 0$
and display the results in Fig.~\ref{fig_delta}.
At high temperatures, the MC data indeed  agree well with the results from 
Eq.~(\ref{eqde}) which is represented by the solid line in Fig.~\ref{fig_delta}.
However, at low temperatures the difference is apparent unless
$E_J/E_0$ is sufficiently small.
In particular, for $E_J \gtrsim E_0$, the MC simulations give the
value approaching unity as the temperature is decreased to zero,  
whereas the value from Eq.~(\ref{eqde}) approaches zero.
This is the case that the second term in Eq.~(\ref{eq_Hcl}) is dominant
over the first term, which leads to the ordering of both $\{ \theta_i \}$ and 
$\{ n_i \}$. This reflects strong correlations between the two, which
have not been taken into account in the SCA.
It is thus concluded that the {\it quantitative} behaviors of the
phase diagrams in Fig.~\ref{fig_selfbd} are accurate only at sufficiently high
temperatures as already shown in  Table~\ref{tab_j}.

\section{Conclusion} \label{sec_conclusion}

We have investigated the quantum phase transitions in two-dimensional
superconducting arrays with various values of external charge frustration
at both zero and finite temperatures. We have considered general capacitance
matrices, allowing both the self- and the junction capacitances.
In the absence of dissipation, the charge variable in the array changes in
discrete quanta of $2e$,
and the phase variable becomes compact in the interval $[-\pi, \pi)$.
We have studied in detail zero-temperature phase transitions 
at various values of the charge density and also considered the system 
at finite temperatures with and without the Josephson energy. 

The states of the system at zero temperature have been determined 
from the investigation of the crossing of the ground-state level and the
first-excited-state level.
At vanishing charge density ($\langle q_i \rangle = 0$),
we have considered the commensurate insulating state, where $q_i = 0$
at all sites, and two types of the incommensurate superconducting
state, i.e., the single-charge (SC) type and the charge-dipole (CD) type.
Unless the self-capacitance vanishes, the SC type
incommensurate state has been found to have lower energy than the CD
type one except at very small values of the charge frustration.
From the comparison of the energies of the
commensurate insulating state and the SC type superconducting
state, computed via the second-order 
perturbation expansions, the phase boundary has been
obtained. The validity of the second-order expansion has been checked
for the array in the self-charging limit, which reveals that the
inclusion of the third-order term does not change significantly the phase diagram.
At finite charge densities, identification of the commensurate
state and of the first-excited incommensurate state is a highly
nontrivial problem. We have thus adopted  the simulated
annealing method and applied  the second-order perturbation theory 
to obtain  a number of lobe-like insulating phases such as
$\langle q_i \rangle = 1/2$, 1/3, 1/4, and 1/5 
in $10 \times 10$ and $12 \times 12$ arrays. As the system size
is increased, more insulating lobes are expected to be observed.
The effects of the junction capacitance have been investigated 
in detail in the insulating phases of $\langle q_i \rangle = 0$ and 1/2:
As $C_1/C_0$ is increased, the insulating region shrinks monotonically for 
$\langle q_i \rangle = 0$, while the insulating region for 
$\langle q_i \rangle = 1/2$ expands with $C_1$ and then shrinks, with its
maximum area at $C_1/C_0 \approx 0.1$.
The MC simulations have displayed that 
the lowest excited state changes as the junction capacitance is increased, 
which in  turn indicates  that the existence of the supersolid phase with
$\langle q_i \rangle = 1/2$ depends on the interaction range of charges: 
The supersolid
phase exists for sufficiently large values of the ratio of the junction
capacitance to the self-capacitance.
In the supersolid phase, it has also been shown that the point defect in the 
charge configuration plays an important role.

At finite temperatures, thermal fluctuations as well as quantum
fluctuations play an important role in the phase transitions of the system.
In the limit of vanishing Josephson coupling strength, we have performed
the MC simulations to obtain $\langle q_i \rangle$
as functions of the charge frustration at various temperatures, and pointed out 
the possibility of the devil's staircase structure in the thermodynamic limit.
As the temperature is increased, higher-order insulating states
disappear but some lower-order ones still remain at 
sufficiently low temperatures. 
In the presence of the Josephson coupling, we have obtained the effective 
classical Hamiltonian
via a variational method in the path integral formalism. A self-consistent
approximation has been applied to the resulting Hamiltonian and phase boundaries 
have been computed for $C_1/C_0 = 0$ and 1.
A reentrant transition at $q_x = 0 $ has been observed for the array
in the self-charging limit ($C_1=0$), while an insulating lobe-like 
structure has been found near $q_x=0.5$ in the case $C_1/C_0=1.0$.
The validity of the formalism used here, the self-consistent approximation (SCA) 
on the effective classical Hamiltonian, has been checked using MC simulations 
in the absence of charge frustration. Although the reentrant behavior of the 
array in the self-charging limit has also been observed in MC simulations,
the SCA appears to be accurate only at sufficiently high temperatures.  

The quantum phase transitions of the superconducting arrays have been studied 
theoretically by many authors via mean-field approximations such as the 
coarse-graining approximation~\cite{bruder} and via the semiclassical methods 
including the variational methods~\cite{bjkim,otterlo} and the WKB 
approximation~\cite{rojas}. Although the mean-field approximations are 
in general expected to give better results at zero temperature, 
where the dimension of the system 
is effectively increased by one, than at finite temperatures, it is well known
that the coarse-graining approximation gives unphysical results even at
zero temperature,  for the arrays without self-capacitances. 
On the other hand, the semiclassical methods 
should be applied with great care at zero temperature, where quantum fluctuations
are strong.
In comparison with these methods, the formalism adopted here to study the 
zero-temperature phase transitions, perturbation expansion combined with  
simulated annealing, appears to be more reliable and can be
applied to other systems systematically. 

%For further studies, it would be interesting to investigate the 
%zero-temperature phase boundaries separating the supersolid phase from the 
%superfluid phase where the commensurability is totally destroyed. 

\begin{acknowledgments}

We thank G.S. Jeon for useful discussions. 
This work was supported in part by the Basic Science Research Institute Program,
Ministry of Education, and in part by the Korea Science and Engineering
Foundation through the SRC Program.

\end{acknowledgments}

\newpage

\newpage

\vskip 3cm

\begin{table}
\caption{Values of $q_e$ in the lowest excited states of the arrays with
size (a) $L=10$ and (b) $L=12$. It is 
shown that $|q_e|$ tends to decrease as $C_1$ is increased.}
\vspace{1cm}
\label{tab_qe}
\begin{center}
{(a) $L=10$ }
\end{center}
\begin{tabular}{r c c c}
$\langle q_i \rangle$ &
       $C_1/C_0 = 0.1$ &
       $C_1/C_0 = 1.0$ &
       $C_1/C_0 = 5.0$ \\ \hline \hline
 1/2 &  -10 & -1 & -1 \\ \hline
 1/5 &  -3 , 4  &  -3 , 3  &  -1 , 1\\
\end{tabular}
\vspace{0.5cm}
\begin{center}
{(b) $L=12$ }
\end{center}
\begin{tabular}{r c c c}
$\langle q_i \rangle$ &
       $C_1/C_0 = 0.1$ &
       $C_1/C_0 = 1.0$ &
       $C_1/C_0 = 5.0$ \\ \hline \hline
 1/2 &  -12 & -1 & -1 \\ \hline
 1/3 &  -8 , 12  &  -4 , 6 &  -2 , 2 \\ \hline
 1/4 &  -4 , 4  &  -4 , 3  &  -2 , 1\\
\end{tabular}
\end{table}

\end{multicols}
\widetext

\begin{table}
\caption{Step width for $\langle q_i \rangle = 0$, 1/3, and 1/2, in 
the $L=12$ array at zero temperature, 
obtained from the data in Fig.~\ref{fig_mc}. The numbers in parentheses
denote the maximum error in the last digits.}

\vspace{1.0cm}

\label{tab_mc}
\begin{tabular}{c|c|c|c}  
$\langle q_i \rangle$ & 
       $C_1/C_0 = 0.1$ &
       $C_1/C_0 = 1.0$ & 
       $C_1/C_0 = 5.0$ \\ \hline \hline
 0 &  0 $\le q_x<$  0.3650(5) & 
      0 $ \le q_x <$ 0.1275(5)  & 
      0 $ \le q_x <$  0.0395(5) \\ \hline
 1/3 &  0.3900(5) $<q_x<$ 0.4055(5) & 
      0.314(1) $ <q_x <$  0.3800(5) & 
      0.317(1)  $ <q_x <$  0.356(1) \\ \hline
 1/2 &  0.4130(5) $<q_x \le$  1/2 & 
      0.4290(5) $ <q_x \le$  1/2 & 
      0.4725(5)  $ <q_x \le$  1/2\\ 
\end{tabular}
\end{table}

\begin{table}
\caption{The critical values of $E_J/E_0$ at $q_x=0$ for $C_1/C_0= 0$ and 
$1.0$. The results from Monte Carlo simulations  and those  
from Figs.~\ref{fig_selfbd} and \ref{fig_1.0bd}
at temperatures $T=0.1$, 0.2, and 0.5 are displayed in the left
and the right columns, respectively. Both results display reentrance for
$C_1/C_0=0$.
The numbers in parentheses denote the maximum error in the last digits.}
\label{tab_j}
\vspace{1.0cm}

\begin{tabular}{c|c|c|c|c }
\mbox{   } $T$ \mbox{   } &  \multicolumn{2}{c|}{$(E_J/E_0)_c$ \ \ for $C_1/C_0=0$}   
    &  \multicolumn{2}{c}{ $(E_J/E_0)_c$ \ \ for $C_1/C_0 =1$} \\
\hline \hline 
\mbox{   } 0.1 \mbox{   }  &  0.642(1)  \ \ & 0.88(1)&  0.158(2) \ \  & 0.28(1) \\
\hline
\mbox{   } 0.2  \mbox{   } &  0.558(2)  \ \  & 0.81(1) & 0.242(4) \ \  & 0.34(1) \\
\hline
\mbox{   } 0.5  \mbox{   } &  0.766(2)  \ \  & 0.91(1)& 0.53(1)  \ \ &  0.57(1) \\
\end{tabular}
\end{table} 
\end{document}